\documentclass[twocolumn, amssymb, nobibnotes, aps, prd]{revtex4-2}

\usepackage{graphicx}% Include figure files
\usepackage{dcolumn}% Align table columns on decimal point
\usepackage{bm}% bold math

\begin{document}

\title{Physics of acceptors in GaN: Koopmans tuned HSE hybrid functional calculations and experiment }% Force line breaks with \\

\author{D. O. Demchenko}
\email{ddemchenko@vcu.edu}
\affiliation {Department of Physics, Virginia Commonwealth University, Richmond, Virginia 23220, USA}

\author{M. Vorobiov}
 \affiliation {Department of Physics, Virginia Commonwealth University, Richmond, Virginia 23220, USA}
\author{O. Andrieiev}
 \affiliation {Department of Physics, Virginia Commonwealth University, Richmond, Virginia 23220, USA}
 \author{M. A. Reshchikov }
 \affiliation {Department of Physics, Virginia Commonwealth University, Richmond, Virginia 23220, USA}
\author{B. McEwen}
 \affiliation {Department of Nanoscale Science and Engineering, SUNY-Albany, Albany, New York 12203, USA}
\author{F. Shahedipour-Sandvik}
 \affiliation {Department of Nanoscale Science and Engineering, SUNY-Albany, Albany, New York 12203, USA}

\date{\today}

\begin{abstract}
The Heyd-Scuseria-Ernzerhof (HSE) hybrid functional has become a widely used tool for theoretical calculations of point defects in semiconductors. It generally offers a satisfactory qualitative description of defect properties, including the donor/acceptor nature of defects, lowest energy charge states, thermodynamic and optical transition levels, Franck-Condon shifts, photoluminescence (PL) band shapes, and carrier capture cross sections. However, there are noticeable quantitative discrepancies in these properties when compared to experimental results. Some of these discrepancies arise from the presence of self-interaction in various parametrizations of the HSE. Other errors are due to the use of the periodic boundary conditions. In this study, we demonstrate that the error corrections scheme based on extrapolation to the dilute limit effectively eliminates the errors due to artificial electrostatic interactions of periodic images and interactions due to the defect state delocalization. This yields parametrizations of HSE that satisfy the generalized Koopmans' condition, essentially eliminating self-interaction from defect state orbitals. We apply this HSE Koopmans tuning individually to a range of cation site acceptors in GaN (Be\textsubscript{Ga}, Mg\textsubscript{Ga}, Zn\textsubscript{Ga}, Ca\textsubscript{Ga}, Cd\textsubscript{Ga}, and Hg\textsubscript{Ga}) and compare the HSE results with experimental data from PL spectra. The Koopmans-compliant HSE calculations show a significantly improved quantitative agreement with the experiment.

\end{abstract}

\maketitle

%\tableofcontents

\section{\label{sec:level1}Introduction}
\begin{figure}
\hspace{-0.28in}
\vspace{-0.1in}
\includegraphics[scale=0.43]{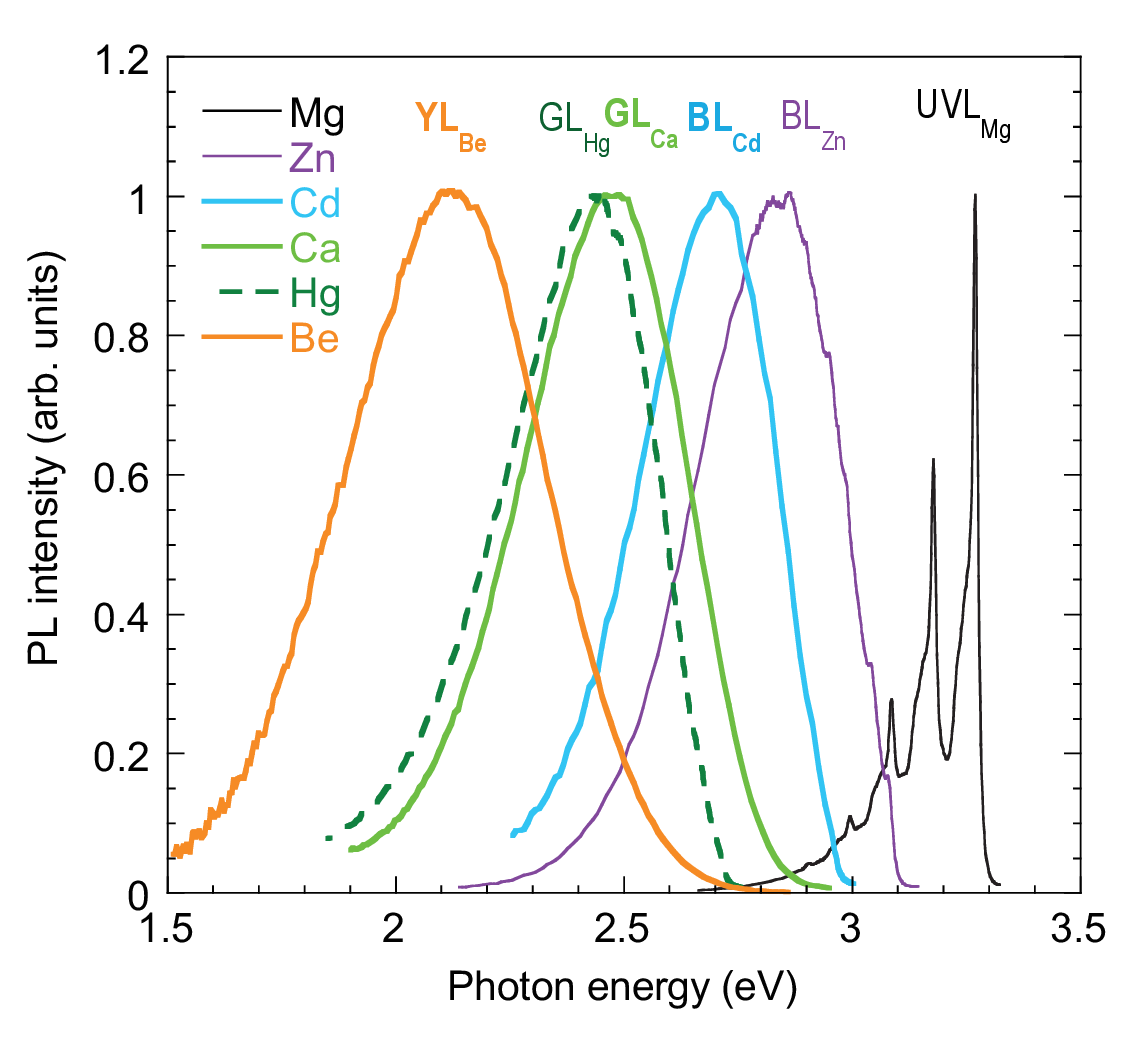}
\caption{\label{fig:All-acceptors} Normalized low-temperature ($T=18$ K) PL spectra due to optical transitions via the cation site acceptors in GaN. PL bands labeled by the respective acceptor are observed in GaN samples doped or implanted by the corresponding impurity.}
\end{figure}

Gallium nitride (GaN) is widespread in a variety of electronics and optical devices, such as light emitting and power electronics devices \cite{Power_electronics,Nakamura_devices}. Improving understanding of the physics of point defects and impurities can lead to improved materials and more efficient devices.
Photoluminescence (PL) experiments on GaN samples grown by doping with various cation site acceptors reveal a series of luminescence bands, shown in Fig. \ref{fig:All-acceptors}, ranging from yellow YL\textsubscript{Be} for the Be\textsubscript{Ga} (Be substituting for Ga atom) acceptor to ultraviolet UVL\textsubscript{Mg} for the Mg\textsubscript{Ga} acceptor. 
The experimental PL spectra yield a range of defect properties, such as energies of the PL maxima, zero phonon lines (ZPL), Franck-Condon (FC) shifts, vibrational energies, and Huang-Rhys factors. 
The fact that all known to date cation site acceptors in GaN are radiative recombination centers presents an opportunity to improve the theory of defects in solids and bring it closer to the experiment. It also offers insights into trends in the behavior of defects in terms of their transition levels and their deep versus shallow nature.

A practical application of acceptors in GaN lies in the development of efficient $n$- and $p$-type materials. While obtaining $n$-type GaN is straightforward, given the existence of various shallow donors, the only acceptor impurity currently capable of reliably producing $p$-type GaN is magnesium substituting for gallium (Mg\textsubscript{Ga}). Additionally, achieving sufficient concentrations of holes (above 10\textsuperscript{18}~cm\textsuperscript{-3}) in Mg-doped GaN remains challenging due to the relatively high acceptor ionization energy (0.22 eV) \cite{Reshchikov-Morkoc, Monemar_Mg, Glaser_Mg}.

A possible alternative to Mg\textsubscript{Ga} acceptor for $p$-type doping is Be\textsubscript{Ga} acceptor, which exhibits similar electronic properties. Despite promising preliminary studies suggesting that the acceptor activation energy of Be\textsubscript{Ga} is approximately half that of Mg\textsubscript{Ga} \cite{Javorek_Be, Teisseyre_Be, Nakano_Be-doping, Demchenko_Be_APL, Demchenko_Be_PRL}, in most cases Be doping produces semi-insulating GaN \cite{Resh_implant-Be, Ben_MOCVD, Resh_MOCVD, Bockiwski_Be-GaN, Storm_Be, Sanches_Be}. Most recent experiments, revised the value of the measured shallow acceptor level of Be\textsubscript{Ga} to a deeper value, as well as confirmed the existence of the deep transition level of the same acceptor \cite{Resh_dual-nature_Be}. 

The dual nature of both Be\textsubscript{Ga} and Mg\textsubscript{Ga} acceptors in GaN, i.e., the coexistence of the shallow (weakly localized) and deep (small polaron) states of the same acceptor, was first predicted theoretically in Ref. \cite{Lany-Zunger_dual-nature}. Recent PL experiments confirmed that there are two states of neutral Be\textsubscript{Ga} acceptor, a deep ground state of the hole at $\sim$0.35 eV above the valence band maximum (VBM) and a metastable shallow state at $\sim$0.2 eV above the VBM \cite{Resh_dual-nature_Be}. 
In contrast to Be\textsubscript{Ga} acceptor, in the case of Mg\textsubscript{Ga}, the theory predicted weakly localized and polaronic acceptor states to have very similar energies \cite{Lany-Zunger_dual-nature, Sun-Zhang_Mg, Demchenko_Mg}. However, to date, no experimental evidence of the dual nature of Mg\textsubscript{Ga} acceptor has been demonstrated, and only the shallow state of this acceptor is observed in PL spectra \cite{Resh_Mg}.

Zn\textsubscript{Ga} acceptor was also theoretically predicted to exhibit two neutral state lattice configurations. However, in this instance, both weakly localized and polaronic states were predicted to be deep \cite{Lany-Zunger_dual-nature}. 
Other theoretical calculations predicted only one of these states, either localized \cite{Droghetti_Zn_Ga, Lyons_acceptors, Lyons_Mg_PRL} or weakly localized \cite{Demchenko_Zn_Ga}. 
The latter theoretical works did not explicitly explore multiple metastable lattice configurations, since locating these deep and shallow states requires numerous calculations relaxing atomic structure starting from different symmetry-breaking defect lattice configurations \cite{Sun-Zhang_Mg}.  
Theoretical predictions of 0/- transition level of Zn\textsubscript{Ga} range from 0.25 to 0.45 eV above the VBM, with the experimental value measured at 0.40 eV \cite{Reshchikov-Morkoc,Demchenko_Zn_Ga}.

Ca\textsubscript{Ga} was theoretically predicted to be a deep acceptor with the 0/- transition level at 1 eV \cite{vdW_Ca_Ga,Lyons_acceptors2} using HSE calculations. Recent calculations using a similar theoretical approach suggested a revised transition level at 0.70 eV, while the PL experiments yielded 0.50 eV \cite{Resh_Ca_Ga}. No weakly localized state could be found theoretically for Ca\textsubscript{Ga} acceptor. 

A comprehensive review of historic experimental data for acceptors in GaN can be found in Ref. \cite{Reshchikov-Morkoc}, while the review of the most recent experimental data for defects in GaN is presented in Ref. \cite{Resh_SPIE}. 
In early studies of GaN samples implanted with Cd or Hg, PL bands with maxima at 2.7 and 2.43 eV were attributed to the Cd\textsubscript{Ga} and Hg\textsubscript{Ga} acceptors, respectively \cite{Pankove}. The ionization energies of these acceptors were estimated as 0.55 and 0.41 eV above the VBM, respectively \cite{Hg_experiment,Monemar_Cd}.
Recent experiments confirmed the attribution of these PL bands and provided more accurate estimates for the acceptor ionization energies, i.e., 0.55 and 0.77 eV above the VBM for  Cd\textsubscript{Ga} and Hg\textsubscript{Ga} acceptors, respectively \cite{Resh_Cd-Hg}. 
Early theoretical works analyzed Cd\textsubscript{Ga} acceptor (as well as several other defects in GaN) using approaches available at the time, such as $\mathbf{k \cdot p}$, Green's function methods, and early local density functional approaches \cite{Vanderbilt, Green_function, k-dot-p}. The calculated 0/- transition level of Cd\textsubscript{Ga} ranged form $\sim$0.2 to 0.65 eV. Hg\textsubscript{Ga} acceptor was not addressed in these early works. 
To the best of our knowledge, there are no recent theoretical studies of Cd\textsubscript{Ga} and Hg\textsubscript{Ga} acceptors in GaN.

\begin{figure}
\hspace{-0.28in}
\vspace{-0.1in}
\includegraphics[scale=0.43]{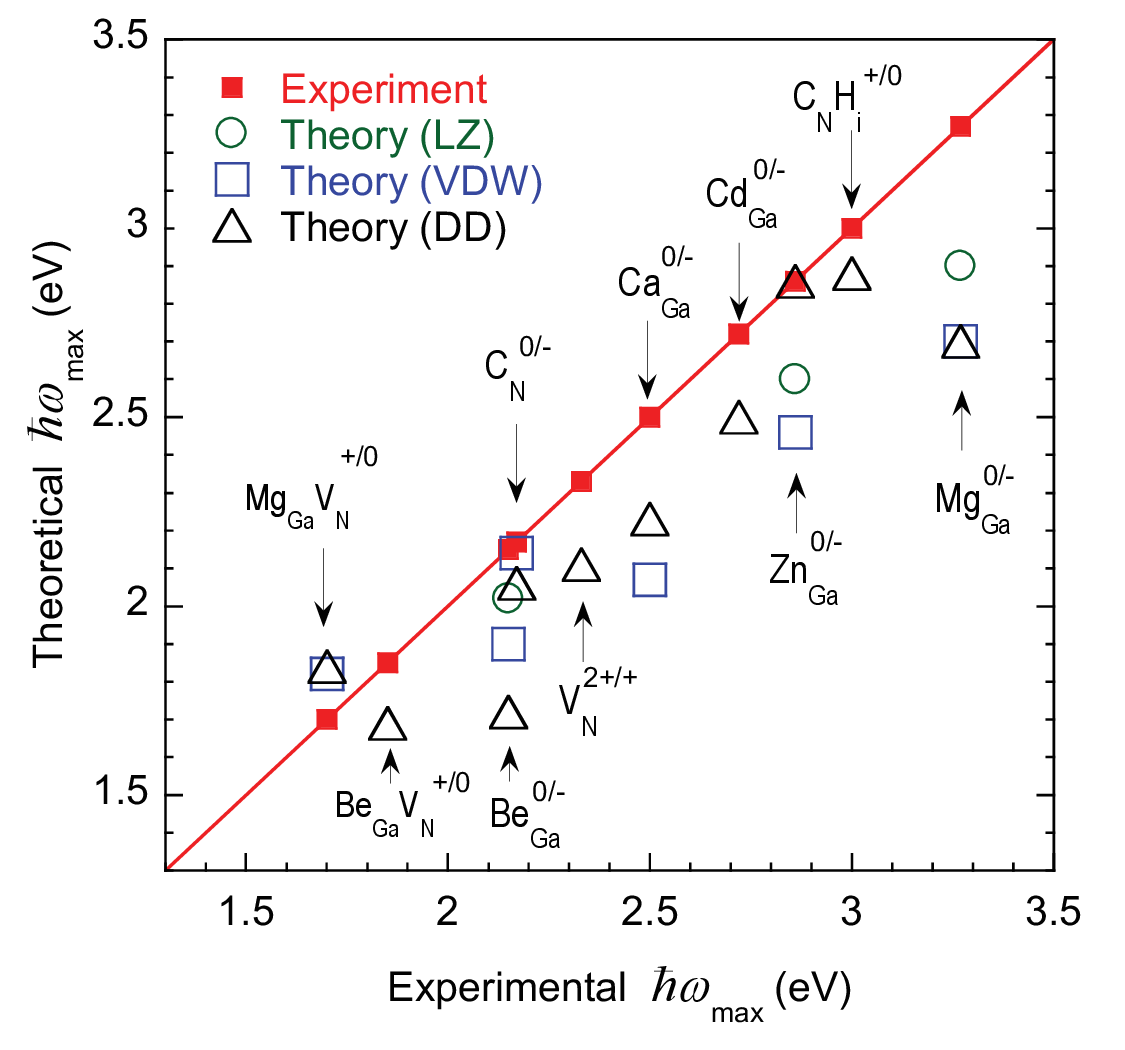}
\caption{\label{fig:theory-vs-exp} Experimental positions of PL band maxima for common defects in GaN compared to recent theoretical calculations.}
\end{figure}

Among theoretical approaches for calculating point defects in semiconductors, HSE hybrid functional has become the most widespread in recent years \cite{Freysoldt_review}. When applied along with the appropriate finite supercell size corrections, it usually produces results in reasonable agreement with the experiment. 
Figure \ref{fig:theory-vs-exp} shows the summary of recent theoretical calculations of PL maxima of different defects in GaN by three research groups (Lany and Zunger (LZ) \cite{Lany-Zunger_dual-nature}, Van de Walle (VDW) \cite{Lyons_acceptors,Lyons_acceptors2}, and D. Demchenko (DD) \cite{Resh_CH-complex,Resh_green, Vorobiov_Be,Resh_Ca_Ga,Resh_Mg,Demchenko_Zn_Ga}) compared with experiment. Overall agreement between theory and experiment is fair, with the average error in theoretical calculations around 0.4 eV. However, there is an overall red-shift in theoretical PL maxima for all defects, indicating a potential systematic error in calculations.  

In this paper, we use an extrapolation procedure for total energy corrections to improve the accuracy of calculated optical properties of defects in semiconductors. In addition to the electrostatic interactions correction, this approach takes into account the delocalization error of defect orbitals during the HSE functional tuning procedure. This is in contrast with previous theoretical works where Koopmans tuning of HSE was performed, taking into account only the electrostatic correction for charged defects \cite{LZ_Koopmans, Lany_Koopmans, Pasquarello_Koopmans, Deak_Koopmans, Demchenko_C_N-Koopmans}. Extrapolations to infinitely large supercells show that: 1) in cases where defect states are weakly localized, nominally neutral defects display electrostatic interaction errors; 2) in some cases, although the defect states appear strongly localized, total energies of defects can be affected by delocalization errors in the amounts comparable with the electrostatic errors. The overlap of the defect state wavefunctions among the periodic images leads to an artificial increase in the energy of the occupied defect state. This energy increase partially cancels the negative electrostatic error, resulting in overall lower values of total energy correction. 
We apply this approach to Koopmans tuning of the HSE hybrid functional separately for cation site acceptors in GaN. These HSE parametrizations are then applied to calculations of optical transition energies, configuration coordinate diagrams, and PL band shapes of these acceptors. 
Comparison with the experiment demonstrates significant improvement in the HSE results. 

\section{\label{sec:methods_level1}Methods}
\subsection{\label{sec:comput_level2} Computational details}
Theoretical calculations were performed using the HSE hybrid functional \cite{HSE} and projector augmented wave (PAW) pseudopotentials \cite{Blochl_PP, Kresse_PP} as implemented in VASP code \cite{VASP1, VASP2}. GW-compatible (fitted to reproduce higher energy scattering properties) PAW pseudopotentials were used throughout this work.
The following atomic shells were treated as valence electrons: 4$s$ and 4$p$ for Ga atom, 2$s$ and 2$p$ for N atom, 2$s$ for Be atom, 3$s$ for Mg atom, 3$d$ and 4$s$ for Zn atom, 3$s$, 3$p$, and 4$s$ for Ca atom, 4$d$ and 4$s$ for Cd atom, 5$p$, 5$d$, and 6$s$ for Hg atom. 
The remaining lower energy electrons were treated as core electrons. 
The HSE was tuned to fulfill the generalized Koopmans' condition for each acceptor separately. The tuning procedure and the appropriate corrections are described in Sec.~II(B). 
Different fractions of exact exchange for each acceptor result in different values of the calculated bandgap. For most acceptors, it is underestimated by a few tenths of an eV. 
Therefore to compare calculated optical properties with the experiment we use the experimental value of the bandgap ($E_g^{exp}=$ 3.50~eV) in calculations of optical transitions. For example, if the thermodynamic transition level of an acceptor obtained from HSE is $E(0/-)$ above the VBM, the ZPL of the PL band is calculated as $E_0=E_g^{exp}-E(0/-)$. The PL maximum is, therefore, $\hbar \omega_{max} = E_0-\Delta_{FC}$, where $\Delta_{FC}$ is the calculated Franck-Condon shift, i.e. lattice relaxation energy following the transition at the PL maximum. Calculations were performed in 300-atom hexagonal (wurtzite) supercells at the $\Gamma$-point, with plane-wave energy cutoffs of 500 eV. Defect atomic structures were relaxed within HSE to minimize forces to 0.05 eV/Å or less. Spurious electrostatic interactions and delocalization errors in calculated total energies were corrected using the extrapolations to the infinitely large supercells as described in Sec. II(B). 
Adiabatic potentials used to plot the configuration coordinate diagrams were obtained by fitting into HSE computed transition energies using the harmonic approximation to map the atomic displacements $\Delta R_i$ onto the configuration coordinate $Q$ following Ref. \cite{Alkauskas_capture}.

\subsection{\label{sec:Koopmans_level2} Koopmans tuning HSE}
In practical applications to various problems in the solid state, the fraction of exact exchange $\alpha$ and the exchange range separation parameter $\mu$ in HSE hybrid functional are often treated as adjustable parameters. There are several commonly used HSE-based approaches to defect calculations in semiconductors. 
They differ by which desired property is targeted when adjusting these parameters, i.e.: 1) reproduce the experimental bandgap \cite{Demchenko_Zn_Ga,Lyons_acceptors,Lyons_acceptors2,Sun-Zhang_Mg}; 2) reproduce the correct behavior of the dielectric function of the host material \cite{Pasquarello_Koopmans}; 3) enforce the generalized Koopmans' condition for defect orbitals \cite{Deak_Koopmans,Demchenko_Mg,Demchenko_C_N-Koopmans}. 
The latter is expected to eliminate self-interaction energy from defect state wavefunctions, albeit at the cost of an error in the calculated bandgap. 
In this work we tune HSE to fulfill the generalized Koopmans' condition following the procedure outlined in Ref. \cite{Lany_Koopmans}. For a fixed defect structure, the amount of exact exchange in HSE is adjusted until the electron addition energy $E(N+1)-E(N)$ is equal to the unoccupied Kohn-Sham eigenvalue $e_i(N)$ of orbital $i$ for a neutral defect with $N$ electrons. In principle, either $\alpha$ or $\mu$ can be varied to adjust the amount of exchange energy in the HSE functional. In Ref. \cite{Deak_Koopmans} it was suggested that adjusting both parameters can result in an HSE parametrization that is both Koopmans compliant and reproduces the correct bandgap. In the above reference, the Ga atom pseudopotentials included $3d$ electrons in the valence shell, which significantly increases computational cost. 
As we show in this work, reproducing the experimental bandgap is less important for capturing the physics of acceptors than enforcing Koopmans' condition. Here we choose to adjust the fraction of exact exchange $\alpha$ while keeping the effective exact exchange screening parameter at a standard $\mu=0.2$ \AA$^{-1}$. 
For acceptors considered here, the Koopmans' condition is fulfilled when the total energy difference between the negative and neutral charge states in the same lattice geometry (that of the neutral acceptor) is equal to the unoccupied defect state eigenvalue, i.e. that of the hole localized on a neutral acceptor.

\begin{figure}
\hspace{-0.28in}
\vspace{-0.1in}
\includegraphics[scale=0.3]{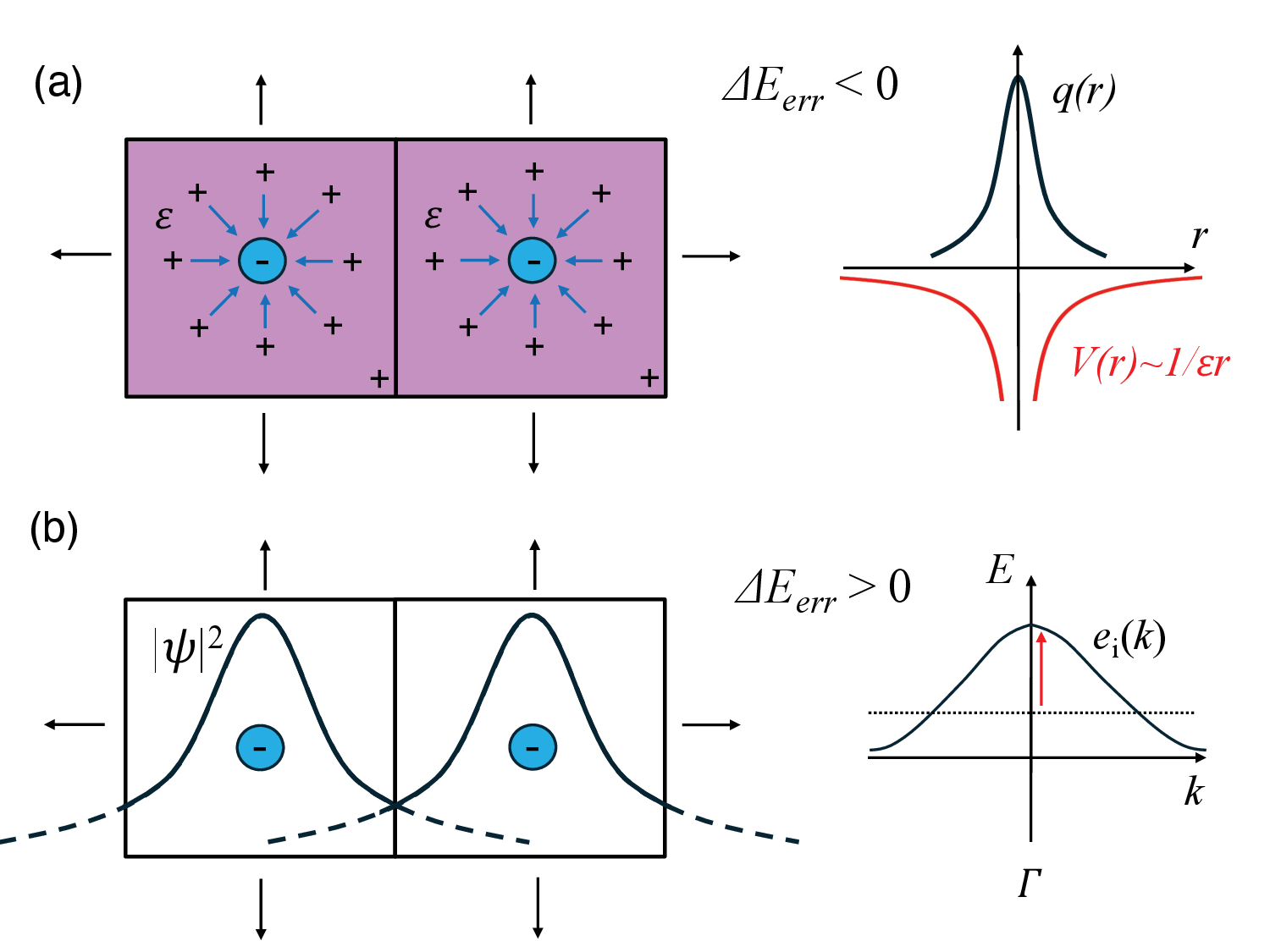}
\caption{\label{fig:Cartoon} Schematics of the two sources of error in the total energy $\Delta E_{err}$ due to periodic boundary conditions. (a) Artificial electrostatic interactions between the periodic images of a charged defect, screened in the medium with the dielectric constant $\varepsilon$, and immersed in the compensating uniform charge. (b) Delocalization error due to the overlap of the occupied defect state wavefunctions between the adjacent supercell images.}
\end{figure}

The problem with the straightforward application of this procedure is that the defect total energies must be corrected for errors due to the periodic boundary conditions. 
The two main sources of error in the calculated energies are illustrated in Fig. \ref{fig:Cartoon}. 
The principal error is electrostatic interactions among periodic images of charged defects (Fig. \ref{fig:Cartoon}(a)). This error in the total energy is negative due to the interactions between the (screened) image charges and compensating background charge density. 
The electrostatic errors are typically corrected using one of the several widely accepted correction schemes \cite{FNV_1,FNV_2,LZ_corr,Oba_corr,Falletta_corr}. In most approaches, the energy correction is calculated from the averaged \textit{ab initio} potential which is matched to the long-range electrostatic potential $V(r)$ from a model charge density $q(r)$. 
It is also suggested that a high-frequency (ion clamped) dielectric constant $\varepsilon_{\infty}$ should be used for unrelaxed defects, and a low-frequency (relaxed-ion) dielectric constant $\varepsilon_0$ should be used when a charged defect is relaxed and ionic screening is significant \cite{Komsa_corr}. It is also typically assumed that energies of neutral defects do not need to be corrected, unless the defect state is delocalized, and a neutral defect resembles an ion in a compensating background \cite{Oba_ZnO}. 
These correction schemes work reasonably well for transition energy calculations. However, there are additional sources of error that must be addressed for HSE tuning, since relatively small errors in calculated total energies of negatively charged $E(N+1)$ and neutral $E(N)$ acceptors can lead to large variations in the optimal amount of exact exchange in HSE \cite{Demchenko_C_N-Koopmans}.

Another significant source of error in the calculated total energy is due to the overlap of the occupied defect state orbitals between the supercell periodic images (Fig. \ref{fig:Cartoon}(b)). 
This error in the total energy is positive because the wavefunction overlap leads to the widening of the impurity band which raises the occupied defect state eigenvalue of an acceptor at the $\Gamma$-point, artificially increasing the calculated total energy. In addition, there can be multiple occupied defect-induced states which can contribute to this error. To our knowledge, there is no model correction scheme for this error in the literature. This error can be partially remedied by shifting the calculations off the $\Gamma$-point to a specially chosen \textbf{k}-point, a mean-value point in the Brillouin zone \cite{Baldereschi}, which should represent a good approximation to the
average value of the defect state. This is a good alternative to a calculation on a regular \textbf{k}-point mesh \cite{Alkauskas_capture}. However, a regular \textbf{k}-point mesh calculation will contain this error as well if the wavefunction overlap is not negligible. 
As we show below, even in cases where defect state wavefunction appears well localized, there can be a significant positive artificial contribution to the computed total energies. 

Since the two errors have opposite signs, taking into account electrostatic corrections only would lead to an over-correction, i.e. overestimated charged defect energies, and therefore acceptor transition levels that are too deep, and calculated PL bands erroneously red-shifted. 
In order to calculate corrections to the defect total energies containing both contributions, here we use an extrapolation scheme, where a series of increasing supercell sizes are used to extrapolate the total energies to an infinitely large supercell \cite{Lambrecht_scaling}. 

\begin{figure}
\hspace{-0.28in}
\vspace{-0.1in}
\includegraphics[scale=0.31]{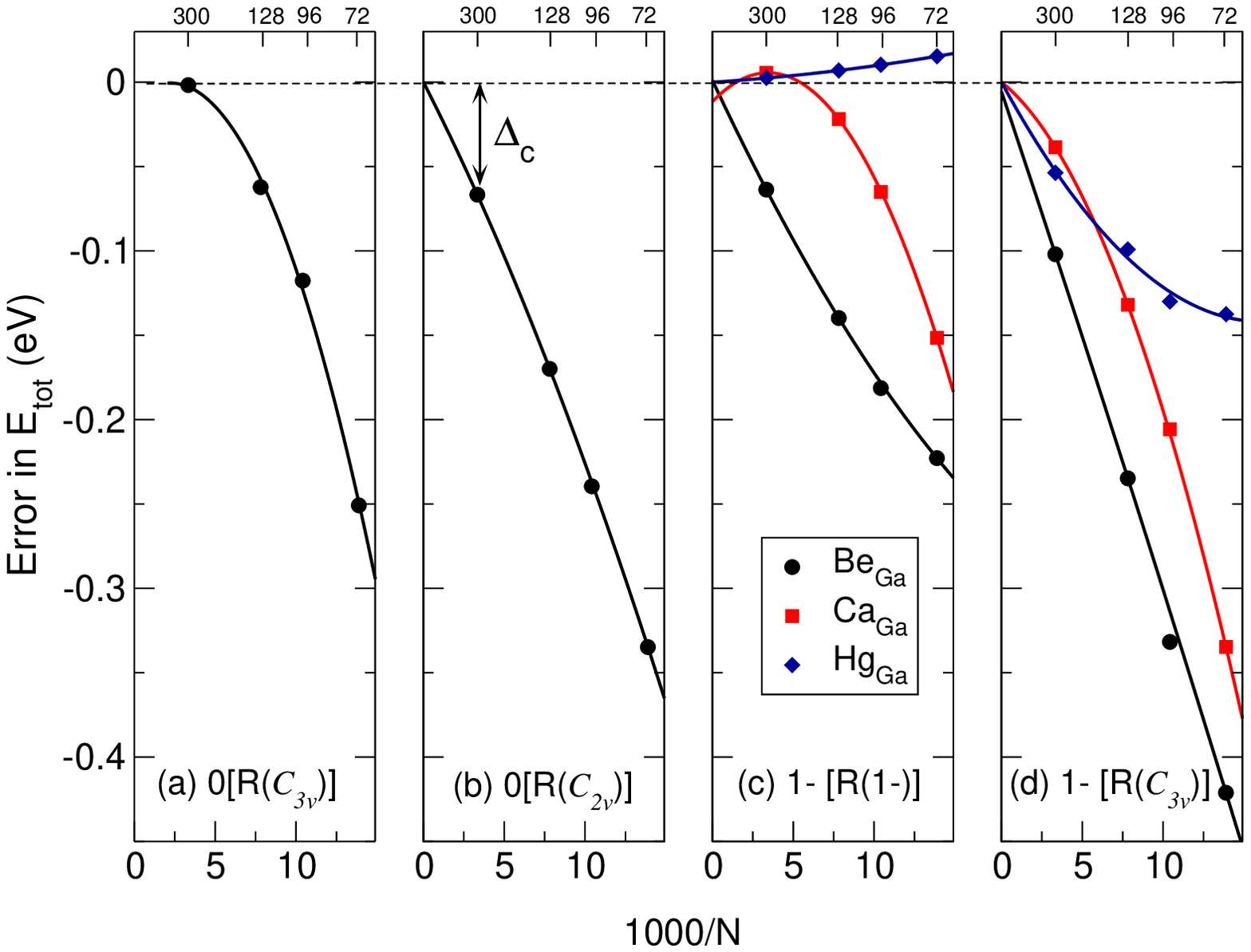}
\caption{\label{fig:Fits} Typical size scaling of the error in defect total energies as a function of the inverse supercell volume in units of 1000/N, where N is the number of atoms in the supercell. The total energies are plotted with respect to the bulk GaN of the same supercell size and extrapolated to the infinite size. Scaling of the (a) neutral Be\textsubscript{Ga} acceptor $C_{3v}$ state; (b) neutral Be\textsubscript{Ga} acceptor $C_{2v}$ state; (c) negatively charged acceptors in their relaxed lattice geometry; and (d) negatively charged acceptors in the neutral $C_{3v}$ lattice geometry. The error correction for a 300-atom supercell is shown as an arrow labeled $\Delta_C$.}
\end{figure}

Figure \ref{fig:Fits} shows the size scaling of the error in total energies for several acceptors in GaN considered in this work. 
The error in defect total energy for each supercell size is obtained from the extrapolation of $E_{defect}-E_{bulk}$ as a function of the inverse number of atoms to an infinitely large supercell.
The delocalization error in total energy scales inversely with the volume of the supercell \cite{vdW_shallow}, leading to linear scaling as a function of $1/N$. The electrostatic image interaction error has two contributions, scaling as $1/L$ and $1/L^3$ (the latter being equivalent to $1/N$) \cite{Komsa_corr}, where $L$ is the size of the supercell, which leads to some deviation from a linear dependence in certain cases. 
To account for this nonlinearity, second-degree polynomials are used to extrapolate the energies to an infinitely large supercell.
The total energies of defects in Fig. \ref{fig:Fits} were computed for HSE parameters $\alpha$=0.3 and $\mu$=0.2 \AA\textsuperscript{-1}. The energy scaling is virtually independent of the HSE parameters, as long as the defect state wavefunction does not substantially change its character. Therefore error corrections calculated for one HSE parametrization, as a part of the tuning procedure, can be used for other HSE parameters for a given charge state of defect. 

\begin{figure}
\hspace{-0.28in}
\vspace{-0.1in}
\includegraphics[scale=0.31]{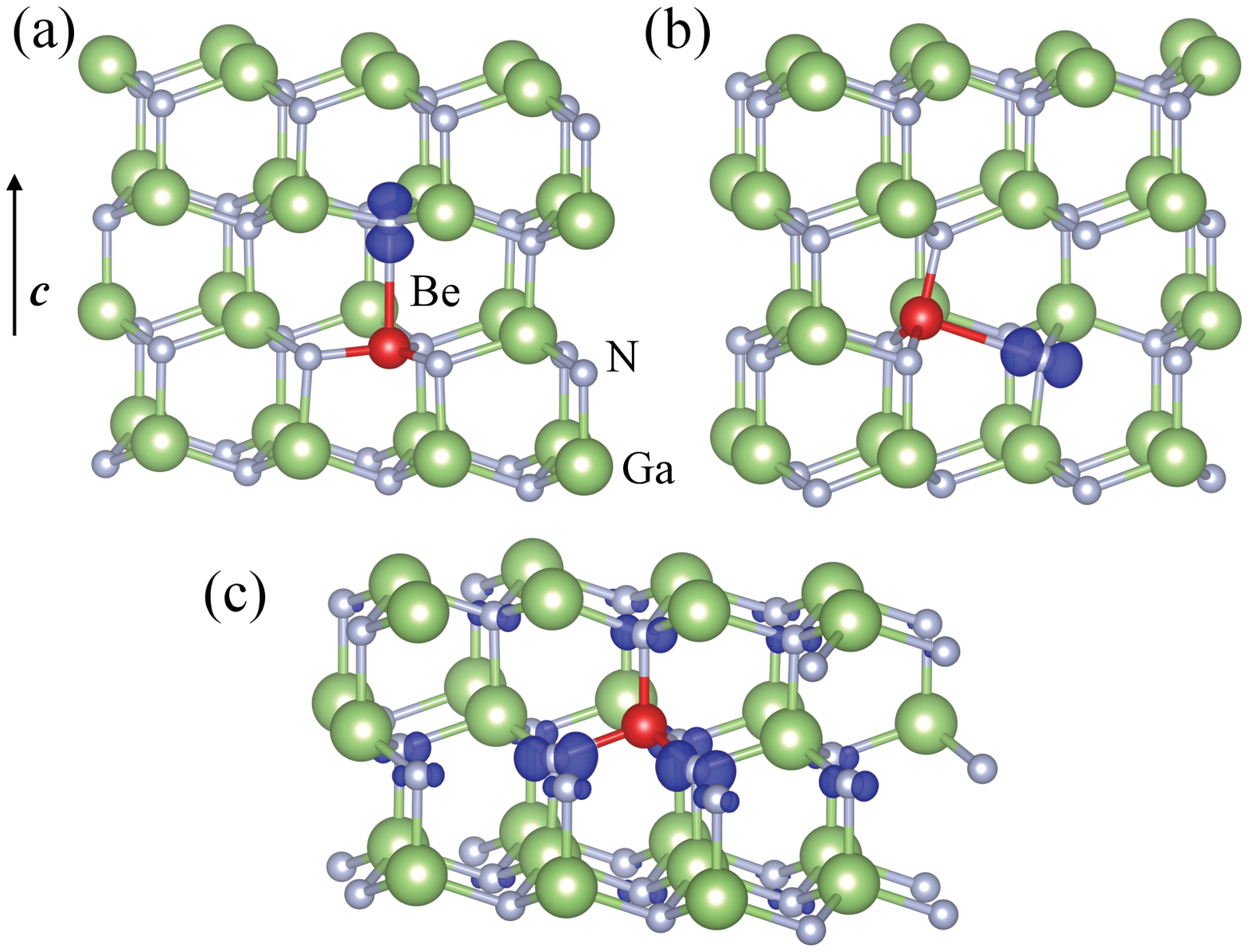}
\caption{\label{fig:WF} Isosurfaces of the spin density of the hole localized on neutral Be\textsubscript{Ga} acceptor at 15\% of the maximum value. (a) Deep polaronic $C_{3v}$ state; (b) deep polaronic pseudo-$C_{3v}$ state; (c) shallow $C_{2v}$ state.}
\end{figure}

In Fig. \ref{fig:Fits}, Be\textsubscript{Ga} shows a typical behavior of a dual nature acceptor in GaN. The two neutral states are labeled 0[R(${C_{3v}}$)] for the deep polaronic state and 0[R(${C_{2v}}$)] for the anisotropically delocalized state, in the literature often referred to as shallow. Here $q$[R($x$)] denotes a charge state $q$ of a defect in a lattice configuration $x$. These states are shown in Fig. \ref{fig:WF}, where the state with $C_{3v}$ symmetry results from the distortion of the Be-N bond along the wurtzite $c$-axis (Fig. \ref{fig:WF}(a)). A similar polaronic state is the result of the distortion of one of the other three in-plane Be-N bonds and is labeled as a pseudo-$C_{3v}$ state \cite{Sun-Zhang_Mg} (Fig. \ref{fig:WF}(b)), since in zinc-blende lattice all four states are equivalent. The shallow state of Be\textsubscript{Ga} acceptor retains the $C_{2v}$ symmetry of the atomic structure, although the hole is anisotropically delocalized in the (0001) wurtzite plane of GaN (Fig. \ref{fig:WF}(c)).

The energy scaling for deep states of Mg\textsubscript{Ga}, Zn\textsubscript{Ga}, and Cd\textsubscript{Ga} is similar to those of Be\textsubscript{Ga} (Fig. \ref{fig:Fits}(a)). Typically we do not expect a significant size scaling from a neutral defect unless the defect state orbital extends beyond the supercell boundaries. 
In case the polaronic $C_{3v}$ (Fig. \ref{fig:Fits}(a)) state of Be\textsubscript{Ga}, the polaron is not fully localized until the supercell size reaches about 300 atoms. 
For smaller supercells, the neutral defect exhibits some periodic image interactions, however, they reduce to zero in a 300-atom supercell. In contrast, the neutral anisotropically delocalized $C_{2v}$ state (Fig. \ref{fig:Fits}(b)) scales virtually linearly with inverse supercell volume, similar to a charged defect. The error correction for a 300-atom supercell is $\Delta_C$ for this nominally neutral state is similar to that of the negatively charged Be\textsubscript{Ga} acceptor (Fig. \ref{fig:Fits}(c)), $\sim$0.07 eV. 

This value of error correction for a negatively charged acceptor in GaN is substantially lower than the values of $\sim$0.25 eV obtained from the more commonly used methods, such as those in Refs. \cite{FNV_1,Falletta_corr}. The reason for this is the partial electrostatic and delocalization error cancellation. In cases of Ca\textsubscript{Ga} and Hg\textsubscript{Ga} acceptors, this error cancellation is even more pronounced (Fig. \ref{fig:Fits}(c)). Surprisingly, in 300-atom supercells, this cancellation is nearly complete, eliminating the error in the total energy of Ca\textsubscript{Ga} and Hg\textsubscript{Ga} negatively charged acceptors in their relaxed lattices (1-[R(1-)]). 
The errors of negatively charged acceptors in the lattice geometry of the neutral defect (1-[R($C_{3v}$)]) are the largest (Fig. \ref{fig:Fits}(d)), due to the absence of ionic screening of the defect charge density. 
Nevertheless, due to a partial error cancellation, these errors are still significantly lower than those obtained by common electrostatic error correction schemes, and vary around 0.05-0.1 eV. Thus, energy scaling suggests that even for defect states that appear fully localized, such as those of the small polarons, some degree of delocalization can still be present. 
The delocalization leads to electrostatic errors in nominally neutral supercells and to partial error cancellation in charged defect supercells due to competing negative electrostatic error and positive wavefunction overlap. 
If delocalization effects are not taken into account and only electrostatic corrections are made during HSE tuning, the $E(N+1)$ is overestimated and the resulting parametrization of HSE functional contains an excess of the exact exchange energy. This leads to the overlocalization of defect state wavefunctions, deeper transition levels, larger lattice relaxation energies (FC shifts), and red-shifted PL maxima. 

\begin{table}[b]
\caption{\label{Table:HSE-tunings}%
Fraction of exact exchange $\alpha$ in HSE functional that fulfills the generalized Koopmans' condition for each acceptor in GaN. The range separation parameter $\mu$ is kept at a standard 0.2 \AA\textsuperscript{-1} ~value. The values of the bulk GaN bandgap calculated using the $\alpha$ ($C_{3v}$) are also provided.}
\begin{ruledtabular}
\begin{tabular}{c c c c}
Acceptor &
$\alpha$ ($C_{3v}$) & $\alpha$ ($C_{2v}$)& $E_g (eV)$\\
\colrule
Be\textsubscript{Ga} & 0.226 & 0.270 & 3.01 \\
Mg\textsubscript{Ga} & 0.274 & 0.270 & 3.26 \\
Zn\textsubscript{Ga} & 0.245 & 0.275 & 3.12 \\
Ca\textsubscript{Ga} & 0.223 & - & 3.00 \\
Cd\textsubscript{Ga} & 0.246 & - & 3.12 \\
Hg\textsubscript{Ga} & 0.222 & - & 3.00 \\
\end{tabular}
\end{ruledtabular}
\end{table}

Table \ref{Table:HSE-tunings} shows parameters $\alpha$ of the HSE functional obtained for each acceptor using the tuning procedure outlined above. Overall all values are below the commonly used $\alpha=0.31$ which reproduces the experimental bandgap of GaN. For dual nature acceptors, two different defect configurations with $C_{3v}$ and $C_{2v}$ symmetries satisfy the Koopmans' condition for two different values of $\alpha$. 
%In cases of Be\textsubscript{Ga} and Mg\textsubscript{Ga} acceptors, the significant delocalization of the shallow $C_{2v}$ states presents an additional complication. On one hand, these states are not localized enough for Koopmans' condition tuning outlined above, since in a delocalized state the self-interaction energy is zero. On the other hand, they are not fully delocalized to be treated as truly shallow defects, for example as in Ref. \cite{vdW_shallow}. For these states, the Koopmans' condition should be fulfilled in an infinitely large supercell, where the defect state wavefunction is fully reproduced. For this reason, a series of calculations was performed, for several $\alpha$ parameters and in several supercell sizes. Using similar extrapolations as in Fig. \ref{fig:Fits}, HSE parameter $\alpha$ were found for shallow states of Be\textsubscript{Ga} and Mg\textsubscript{Ga}, for which the electron addition energy is equal to the defect state eigenvalue in an infinitely large supercell. 
Overall, the values of $\alpha$ between 0.22 and 0.27 provide reasonable results for all acceptors considered here. Therefore a rough estimation of the defect properties can be made using a standard 0.25 fraction of exact exchange in HSE. Then, more accurate results can be obtained with a Koopmans compliant HSE, noting the orbital dependence, where two different states of the same acceptor require two different $\alpha$ parameters. Below, in Sec. III - V we compare the theoretical results obtained with the above parametrizations of HSE with the experiment.

\subsection{\label{sec:exp_level2} Experimental methods}

\subsubsection{Samples}
Samples analyzed in this work are GaN layers with thicknesses between 1 and 10 $\mu$m grown on c-plane sapphire substrates. The Mg- and Zn-doped GaN samples were grown by the hydride vapor phase epitaxy (HVPE) method. The details about these samples can be found in Refs. \cite{Resh_Mg} and \cite{Resh_abrupt-tunable}. Note that the UVL\textsubscript{Mg} and BL\textsubscript{Zn} bands are often strong in undoped GaN due to contamination with Mg and Zn impurities, thanks to large hole-capture coefficients for the Mg\textsubscript{Ga} and Zn\textsubscript{Ga} acceptors. The Be-doped GaN samples were grown by metalorganic chemical vapor deposition (MOCVD) and molecular beam epitaxy (MBE) methods \cite{Resh_dual-nature_Be}. Ion implantation was the primary source for other cation-site acceptors. For this, undoped or Si-doped GaN layers grown by HVPE were capped with a thin (40-70 nm) AlN layer and implanted with \textsuperscript{40}Ca\textsuperscript{+}, \textsuperscript{114}Cd\textsuperscript{+}, and \textsuperscript{202}Hg\textsuperscript{+} ions at $T$ = 500-600 $^{\circ}$C at the CuttingEdge Ions, LLC. The concentration of implanted impurities was about 10\textsuperscript{17} cm\textsuperscript{-3} in the near-surface 100 nm-thick layer of GaN, which is the optimal depth for PL experiments. After the implantation, the semiconductor wafers were cut into 5$\times$5 mm squares and annealed in nitrogen ambient at 1100 $^{\circ}$C or 1200 $^{\circ}$C for one hour. The AlN layer was removed by etching before PL measurements. Further details about the samples implanted with Ca, Cd, and Hg ions can be found in Refs. \cite{Resh_Ca_Ga, Resh_Cd-Hg}.

\subsubsection{PL measurements}
Steady-state PL was excited with a HeCd laser, dispersed by a 1200 rules/mm grating in a 0.3-m monochromator, and detected by a cooled photomultiplier tube. A closed-cycle optical cryostat was used for temperatures between 18 and 320 K. The as-measured PL spectra were corrected for the measurement system's spectral response. The PL intensity was additionally multiplied by $\lambda^3$, where $\lambda$ is the light wavelength, to present the PL spectra in units proportional to the number of emitted photons as a function of photon energy \cite{Resh_two-state_C_N}. Other details about PL measurements can be found in Ref. \cite{Resh_measurement}.

\section{Results and Discussion}

\begin{figure}
\hspace{-0.28in}
\vspace{-0.1in}
\includegraphics[scale=0.32]{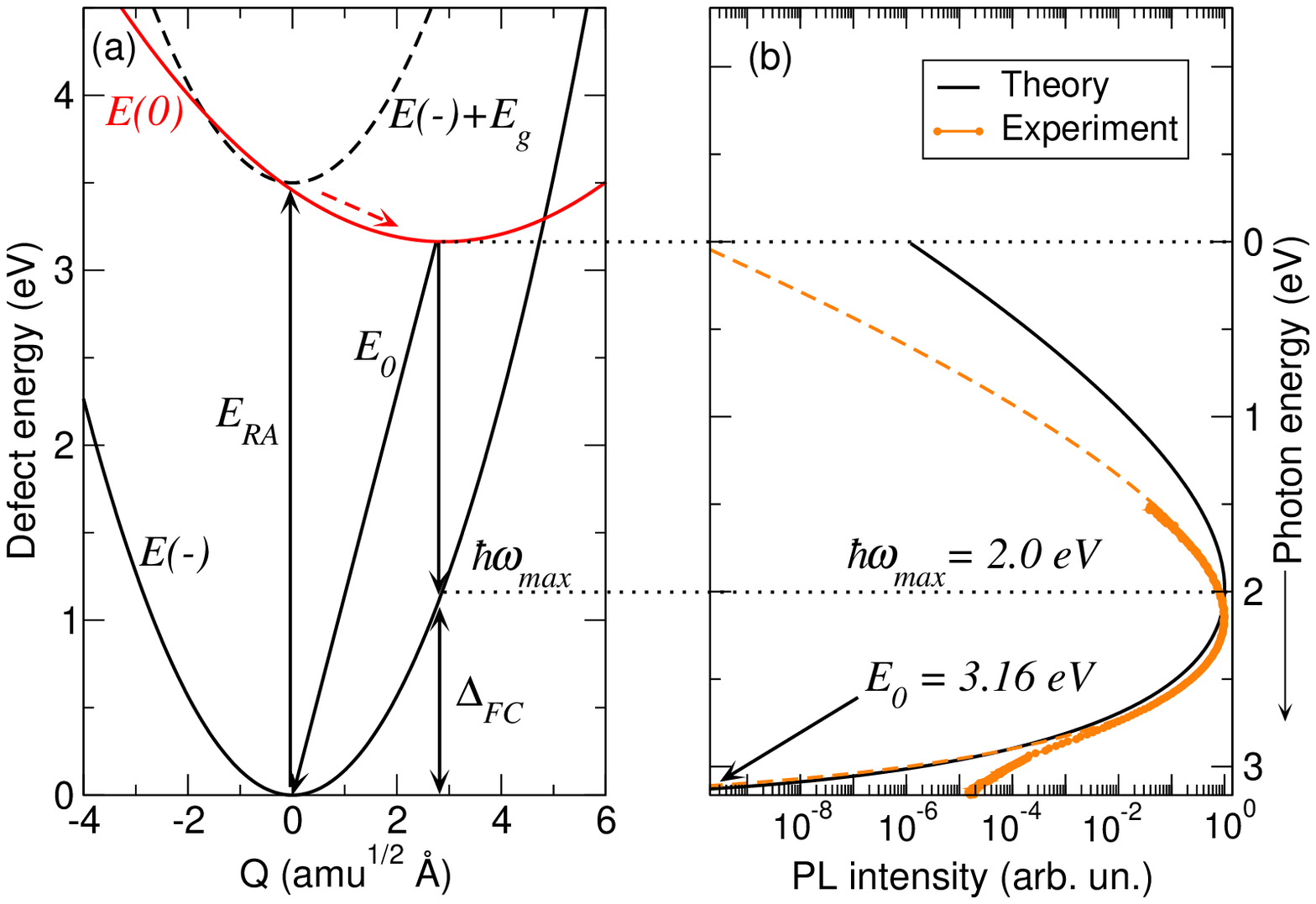}
\caption{\label{fig:Be} Configuration coordinate diagram for the deep polaronic state of Be\textsubscript{Ga} acceptor calculated with the Koopmans tuned HSE (a). $E_{RA}$ is the calculated resonant absorption, $E_0$ is the zero phonon line (ZPL), $\hbar \omega_{max}$ is the energy of the PL maximum, and $\Delta_{FC}$ is the Franck-Condon shift. The diagram is plotted against the calculated PL band shape (b). The solid line is the band shape obtained from Eq. (\ref{Eq:I-PL}) using the HSE computed PL parameters.
The experimental PL spectrum for the YL\textsubscript{Be} band observed in Be-doped GaN is shown with the filled symbols. 
The fit of Eq. (\ref{Eq:I-PL}) into the experiment using the measured $E_0=3.15$ eV and $\hbar \omega_{max} = 2.15$ eV, and a fitting parameter $S_e=20$ is shown with the dashed line.}
\end{figure}

In this section, we present the configuration coordinate diagrams for the optical transitions due to radiative recombination between a free electron and a hole localized on an acceptor. These diagrams are plotted against the PL band shape (in the literature also referred to as the line shape) calculated using the equation for PL intensity $I^{PL}$ derived for the one-dimensional configuration coordinate diagram in Ref. \cite{Resh_green}:
\begin{equation}
I^{PL}\propto\exp \left[ -2 S_e \left(\sqrt{\frac{E_0+0.5\hbar\Omega_e-\hbar\omega}{E_0+0.5\hbar\Omega_e-\hbar\omega_{max}}}-1\right)^2
\right].
\label{Eq:I-PL}
\end{equation}
Here, $S_e$ and $\hbar\Omega_e$ are the Huang-Rhys factor and the vibrational energy in the excited state of the defect (the neutral charge state for an acceptor), $E_0$ is the energy of the ZPL, $\hbar\omega_{max}$ is the energy of the PL maximum. These parameters can be obtained from HSE and the theoretical PL band shape can be plotted. For example, HSE computed adiabatic potentials for the deep state of Be\textsubscript{Ga} acceptor (Fig. \ref{fig:Be}(a)) are used to compute the theoretical PL band shape (solid line in Fig. \ref{fig:Be}(b)). 
The theoretical band shape can be compared to the experimentally measured PL spectrum of the yellow luminescence YL\textsubscript{Be} (filled symbols in Fig. \ref{fig:Be}(b)) in Be-doped GaN samples.
Furthermore, experimental data can also be used to fit into Eq. (\ref{Eq:I-PL}) (dashed line in Fig. \ref{fig:Be}(b)) to obtain experimental values for some of these parameters, where direct observation is not possible. 
The value of $0.5\hbar\Omega_e$ in Eq. (\ref{Eq:I-PL}) is practically negligible. 
Only two parameters ($S_e$ and $E_0$) determine the shape of a PL band with a known maximum at $\hbar\omega_{max}$. 
Further, if $E_0$ is known from independent experiments, such as direct observation of ZPL peak or thermal quenching of PL, only one fitting parameter $S_e$ remains. This approach is adopted hereafter in sections III.A-F, where theoretical PL bands (solid lines) obtained from HSE and using Eq. (\ref{Eq:I-PL}) are compared to experimental spectra (filled symbols) and the fits of experimental data into Eq. (\ref{Eq:I-PL}) (dashed lines). The latter also tests the applicability of the 1-D configuration coordinate model to the problem of optical transitions via defects in semiconductors.  

The parameter $\hbar\Omega_e$ can be found independently from the temperature-related broadening of the PL band's full width at half maximum (FWHM or $W$). For defects with a strong electron-phonon coupling (Gaussian shape of a PL band), the following $W(T)$ dependence is expected \cite{Reshchikov-Morkoc}: 
\begin{equation}
W(T) = W(0) \sqrt{\coth \left(\frac{\hbar\Omega_e}{2kT} \right)}.
\label{Eq:W(T)}
\end{equation}
By fitting the $W(T)$ dependences for the analyzed PL bands with Eq. (\ref{Eq:W(T)}) we have obtained the values of $\hbar\Omega_e$ for acceptors studied here. Below we compare the experimental PL band shapes and measured PL parameters with those obtained from Koopmans tuned HSE.  

\subsection{\label{sec:Be-level2} Be\textsubscript{Ga}}

\subsubsection{Deep $C_{3v}$ state of Be\textsubscript{Ga}}
Figure \ref{fig:Be}(a) shows the configuration coordinate diagram for the deep $C_{3v}$ state of Be\textsubscript{Ga} acceptor. This diagram is plotted against the PL band shape calculated using HSE parameters and Eq. (\ref{Eq:I-PL}) (solid line in Fig. \ref{fig:Be}(b)). The measured YL\textsubscript{Be} band is shown with filled symbols and the fit of Eq. (\ref{Eq:I-PL}) into the experiment is shown with the dashed line in Fig. \ref{fig:Be}(b). 
HSE calculations predict the deep state of Be\textsubscript{Ga}  ($C_{3v}$ state in Fig. \ref{fig:WF}) to be the ground state of this acceptor. The 0/- transition level is calculated to be 0.34 eV above the VBM, corresponding to the ZPL of 3.16 eV, as shown in Fig. \ref{fig:Be}. The polaronic nature of this state is manifested by the large distortion of the Be-N bond to 2.63 \AA\ from the 1.96 \AA\ equilibrium Ga-N bond (see Fig. \ref{fig:WF}). 
This leads to a large shift in configuration coordinate $Q$ between the equilibrium positions of the neutral and negative charge states, about 2.9 amu\textsuperscript{1/2}\AA. As a result, the intensity of the ZPL in the PL spectrum is very low and cannot be observed in the experiment. 
The experimental value of ZPL (3.12 eV) for the YL\textsubscript{Be} band was obtained from the position of the transition level of the deep $C_{3v}$ state (at 0.38 eV above the VBM) \cite{Resh_dual-nature_Be}.
This result is in good agreement with a theoretical value of 3.16 eV. Resonant absorption is calculated to be 3.46 eV, which is close to the band-to-band transition of 3.5 eV. Absorption of a photon above the bandgap creates an electron-hole pair, which relaxes into the respective band edges (i.e., negatively charged acceptor $E(-)$ plus the energy of the bandgap $E_g$, dashed line in Fig. \ref{fig:Be}(a)). 
The configuration coordinate diagram obtained from HSE predicts the nonradiative hole capture by the deep state of Be\textsubscript{Ga} acceptor without a potential barrier, dashed arrow in Fig. \ref{fig:Be}(a), transferring the acceptor into a neutral charge state. 
The calculated PL maximum at 2.0 eV is somewhat red-shifted compared to the experimental value at 2.15 eV (at $T=18$ K) due to an overestimated Franck-Condon shift, 1.16 eV versus the measured 1.0 eV. This is one contribution to the discrepancy between the HSE-predicted and measured PL band shape. As shown in Fig. \ref{fig:Be}(b), the high energy side of the PL band shape is well reproduced. However, there is a discrepancy in the PL band width on the low-energy side. 
The second contribution is the difference in vibronic parameters calculated from HSE and obtained from the experiment. 
For the ground state (negative charge state) of Be\textsubscript{Ga} acceptor HSE yields $S_g=34$ and $\hbar\Omega_g=34$ meV. 
These parameters do not contribute to the PL band shape, as PL intensity in Eq. (\ref{Eq:I-PL}) does not depend on them. From the experiment, only the product of these two parameters can be found, i.e., the measured Franck-Condon shift of 1.0 eV.

The HSE calculated Huang-Rhys factor in the excited (i.e. neutral) state of Be\textsubscript{Ga} acceptor $S_e$ is 17 (i.e. the average number of phonons emitted during relaxation following the resonant absorption), with a corresponding excited state vibronic energy $\hbar \Omega_e =$ 17 meV. 
Reasonable fits of Eq. (\ref{Eq:I-PL}) into the experimental PL band shape can be obtained with the excited state Huang-Rhys factor $S_e$ = 20 $\pm$ 5, close to the theoretical value. However, temperature dependence of PL band width yields $\hbar \Omega_e$ = $38\pm5$ meV \cite{Resh_bockowski}. 
This disagreement indicates that either the 1-D configuration potential model is inadequate and multiple vibrational modes of defect need to be taken into account, or that the single effective mode that reproduces the PL band shape is not directed along the straight line connecting the neutral R(0) and the negative R(1-) charge state configurations in the $Q$-space, as commonly chosen in the literature \cite{Alkauskas_capture, Alkauskas_line-shape, Lyons_acceptors3} as well as this work.

It should be noted that in one of our previous works on Be\textsubscript{Ga} acceptor, we erroneously attributed YL\textsubscript{Be} band to the Be\textsubscript{Ga}-V\textsubscript{N}-Be\textsubscript{Ga} complex \cite{Vorobiov_Be}. One reason for this was a substantial discrepancy between the HSE calculated optical transitions of the isolated Be\textsubscript{Ga} acceptor and those measured for the YL\textsubscript{Be} band. Koopmans tuning of HSE in that work was performed using larger electrostatic corrections, following the commonly used correction scheme \cite{FNV_1, FNV_2}, resulting in a larger exact exchange energy in the HSE functional, and a deeper predicted 0/- transition level of Be\textsubscript{Ga} acceptor. Further experimental and theoretical studies pointed towards the isolated Be\textsubscript{Ga} acceptor as a source of the YL\textsubscript{Be} band in Be-doped GaN \cite{Resh_dual-nature_Be}. Here, the extrapolation-based tuning of HSE also confirms this, producing good agreement between HSE-calculated transitions via the Be\textsubscript{Ga} acceptor and those measured for the YL\textsubscript{Be} band.  

\subsubsection{Shallow $C_{2v}$ state of Be\textsubscript{Ga}}
In recent years there has been some uncertainty regarding the PL produced by the shallow state of Be\textsubscript{Ga} acceptor. The experimental data suggests that an ultraviolet band (labeled UVL\textsubscript{Be}) in some (but not all) Be-doped samples with a sharp peak at 3.38 eV and a series of well-resolved phonon replicas is produced by the shallow state of an acceptor with a transition level at 0.113 eV above the VBM \cite{Demchenko_Be_PRL, Demchenko_Be_APL, Vorobiov_Be,Resh_Be+F}. This band is clearly produced by a Be-related shallow acceptor. For this acceptor, measurements show large hole and electron capture cross-sections (hole and electron capture coefficients are $\sim$10\textsuperscript{-6} and 10\textsuperscript{-11} cm\textsuperscript{3}/s, respectively \cite{Vorobiov_Be,Resh_Be+F}). Therefore a shallow state of Be\textsubscript{Ga} acceptor was a logical attribution of UVL\textsubscript{Be}. 
However, recently a new set of experimental data suggested that the shallow state of Be\textsubscript{Ga} acceptor produces a somewhat deeper 0/- transition level at 0.24 eV above the VBM \cite{Resh_dual-nature_Be}. 
A corresponding different Be\textsubscript{Ga} UVL band emerges with increasing temperature above 140~K and is labeled UVL\textsubscript{Be3}. Solutions to a set of rate equations, describing transitions between pseudo-$C_{3v}$, $C_{3v}$, and $C_{2v}$ states (labeled, Be1, Be2, and Be3 in Ref. \cite{Resh_dual-nature_Be}) indicated that UVL\textsubscript{Be3} is produced by the $C_{2v}$ state of Be\textsubscript{Ga} acceptor, while the most shallow acceptor to date with transition level at 0.113 eV, responsible for the sharp UVL\textsubscript{Be}, remains unidentified.

\begin{figure}
\hspace{-0.28in}
\vspace{-0.1in}
\includegraphics[scale=0.32]{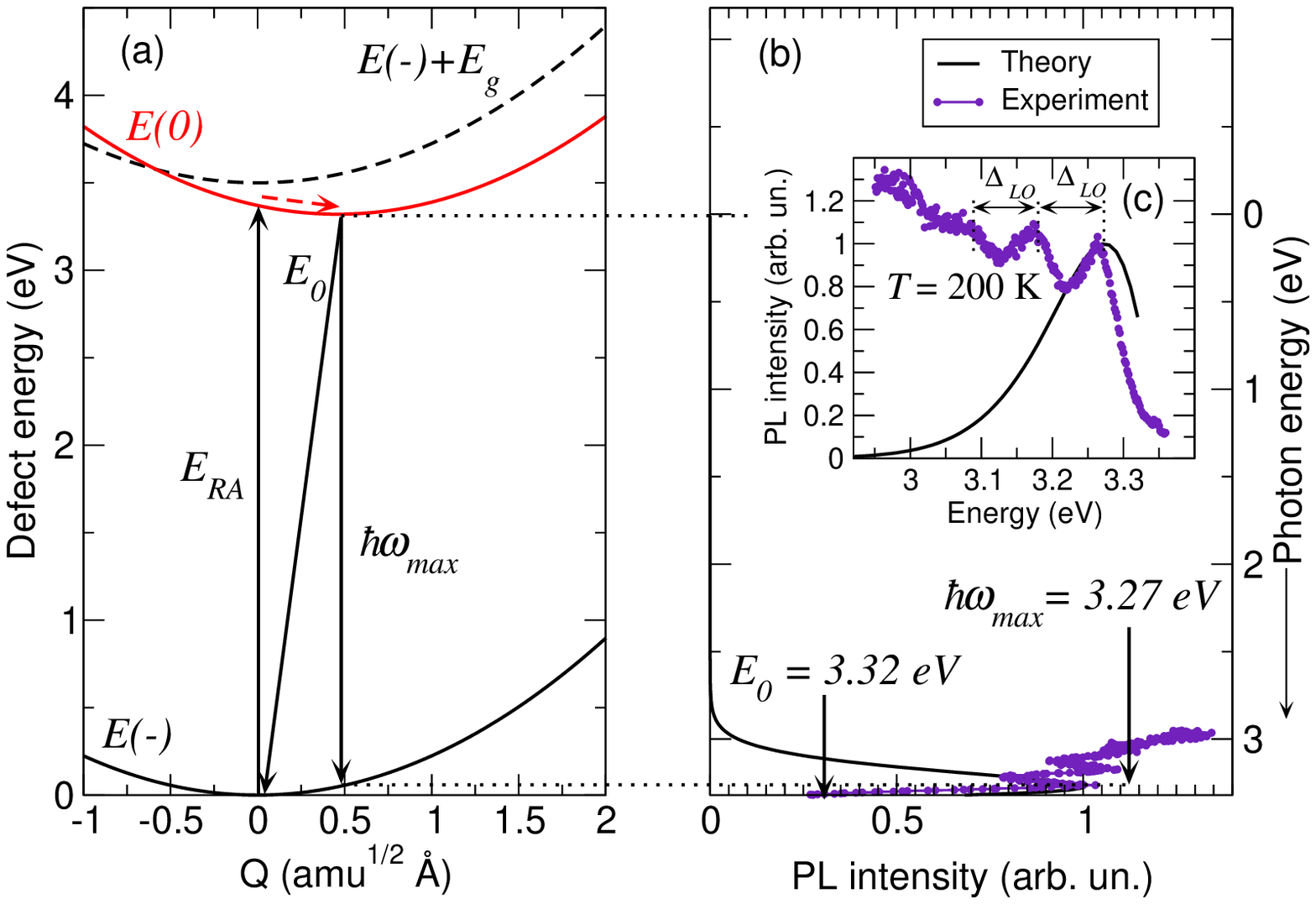}
\caption{\label{fig:Be-shallow} Calculated configuration coordinate diagram for the shallow $C_{2v}$ state of Be\textsubscript{Ga} acceptor (a). The diagram is plotted against the calculated PL band shape and measured UVL\textsubscript{Be3} band (b). The solid line is the band shape obtained using Eq. (\ref{Eq:I-PL}) with HSE computed PL parameters, compared with the experimental PL spectrum (filled symbols) for the UVL\textsubscript{Be3} band in Be-doped GaN measured at $T=$ 200K. The inset (c) shows the details of the experimental (filled symbols) and calculated (solid line) PL band shapes. $\Delta_{LO}$ indicates the phonon replica separation by 92 meV, the energy of the LO phonon mode in bulk GaN.}
\end{figure}

Figure \ref{fig:Be-shallow}(a) shows the calculated configuration coordinate diagram for the shallow $C_{2v}$ state of Be\textsubscript{Ga} acceptor and the PL band shape calculated from this diagram compared with experimental UVL\textsubscript{Be3} band (Fig. \ref{fig:Be-shallow}(b)). The calculated ZPL is 3.32 eV and the PL maximum is 3.27 eV, the inset (Fig. \ref{fig:Be-shallow}(c)) shows a zoom-in of the PL spectrum in this energy range. The calculated and measured transition energies are in good agreement. The experimentally observed ZPL as the high energy peak of the spectrum is at 3.26 eV. The calculated value of the Franck-Condon shift is a very low 0.05 eV.
Since the 1-D configuration coordinate formula for the PL band shape (Eq. (\ref{Eq:I-PL})) was derived as a projection of the excited state $n=0$ vibronic wave function on the ground state adiabatic potential, it does not account for the phonon replicas. At low temperatures, for a shallow acceptor with a low adiabatic potential separation $\Delta Q$, Eq. (\ref{Eq:I-PL}) would reproduce the PL spectrum envelope. This presents a difficulty for comparison with the experimental PL spectrum since UVL\textsubscript{Be3} is not observed at $T<$ 140 K. In Fig. \ref{fig:Be-shallow}(c), UVL\textsubscript{Be3} band is measured at $T=$ 200 K. One notable result of this is increasing intensity of the phonon replicas for decreasing energies, while at low $T$ phonon replicas are expected to have lower intensities. This is due to specifics of the potential landscape of $C_{2v}$ state of Be\textsubscript{Ga} acceptor and population of higher $n>0$ vibrational levels of the adjacent $C_{3v}$ state, which contributes to the spectrum at higher temperatures \cite{Resh_dual-nature_Be}. 

The experimental PL band shows several phonon replicas separated by $\Delta_{LO}$=92 meV, i.e. the energy of the LO phonon mode in bulk GaN. This suggests that, at least for shallow states, the energies of the phonon replicas are determined by the vibrational mode of the defect ground state with stronger coupling to the bulk LO phonon mode.
HSE calculated 1-D configuration coordinate diagram for the mode along the straight line between defect charge state configurations R(0) and R(1-) predicts the ground state vibrational energy of $\hbar\Omega_g$=43 meV with Huang-Rhys factor $S_g$=1. For the excited state, similar values $\hbar\Omega_e$=43 meV and $S_e$=1 are predicted due to similarities of the adiabatic potentials in 1- and 0 charge states for the shallow acceptor. These values were used to calculate the theoretical PL band shape shown in Fig. \ref{fig:Be-shallow}(b). Although this band shape is difficult to compare to the experiment, the defect energetics involved are well reproduced.

\subsection{\label{sec:Mg-level2} Mg\textsubscript{Ga}}

\subsubsection{Deep $C_{3v}$ state of Mg\textsubscript{Ga}}

\begin{figure}
\hspace{-0.28in}
\vspace{-0.1in}
\includegraphics[scale=0.32]{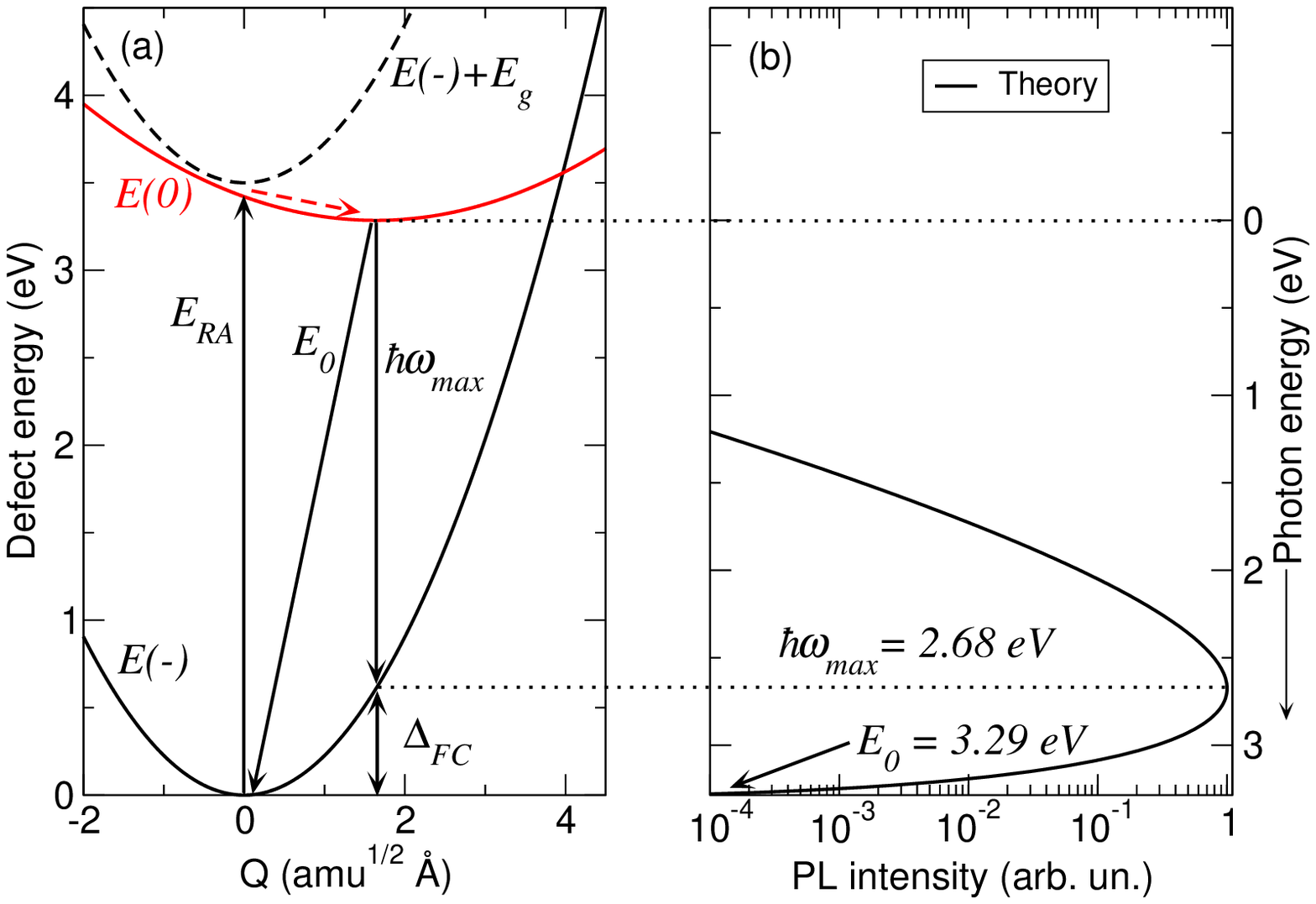}
\caption{\label{fig:Mg} Calculated configuration coordinate diagram for the deep state of Mg\textsubscript{Ga} acceptor (a); corresponding HSE predicted PL band shape (b). This PL band has not been observed in PL experiments.}
\end{figure}

Figure \ref{fig:Mg} shows the HSE calculated configuration coordinate diagram for the deep $C_{3v}$ state of Mg\textsubscript{Ga} acceptor and the corresponding calculated PL band shape. As shown in Fig. \ref{fig:Mg}(a), adiabatic potentials for the negative charge state in the presence of an electron-hole pair (dashed line in Fig. \ref{fig:Mg}(a)) and the neutral charge state do not intersect. This suggests that a nonradiative hole capture path might be blocked. 
However, the calculated energy difference in the adiabatic potentials $E(0)$ and $E(-)+E_g$ is about 0.05 eV, indicating that a slight error in HSE total energies can result in a fitted adiabatic potential which could make this difference vary significantly, therefore this result is inconclusive. 
The photogenerated hole can also transfer to the deep state after being initially captured by the shallow state (see Sec. IV) if there is no significant barrier between these states. 
For the neutral polaronic state of the acceptor, Koopmans tuned HSE predicts a wide blue PL band with a maximum of 2.68 eV and a ZPL at 3.29 eV (Fig. \ref{fig:Mg}(b)). However, this band is not observed in the experiment. There can be several reasons for this. First, as discussed in Ref. \cite{Demchenko_Be_PRL}, the deep polaronic state of the acceptor competes for the photogenerated holes with the shallow state. As a result of the lower hole capture rate and very similar energies of the deep and shallow states, the intensity of this wide PL band is expected to be low and likely below the background. Second, as suggested in Ref. \cite{Resh_dual-nature_Be}, the relative intensities of the PL bands from the deep and shallow states of the same acceptor are also dependent on the potential barrier between the two states. If the hole initially captured by the shallow state, can be transferred to the deep state over a low barrier, then both PL bands can co-exist, as in the case of  Be\textsubscript{Ga} acceptor. If, on the other hand, the barrier is high for the thermal jump from shallow to deep state, only the PL from the shallow state will be observed. Third, if the energies of shallow and deep states are very similar, and if no barrier exists between the two states, then the faster transitions would dominate the PL, i.e., a weakly localized shallow state, which captures both holes and electrons more efficiently, will dominate the PL spectrum. A quantitative discussion of the PL intensity and its dependence on defect energetics and carrier capture rates is given in Sec. IV.

\subsubsection{Shallow $C_{2v}$ state of Mg\textsubscript{Ga}}

\begin{figure}
\hspace{-0.28in}
\vspace{-0.1in}
\includegraphics[scale=0.32]{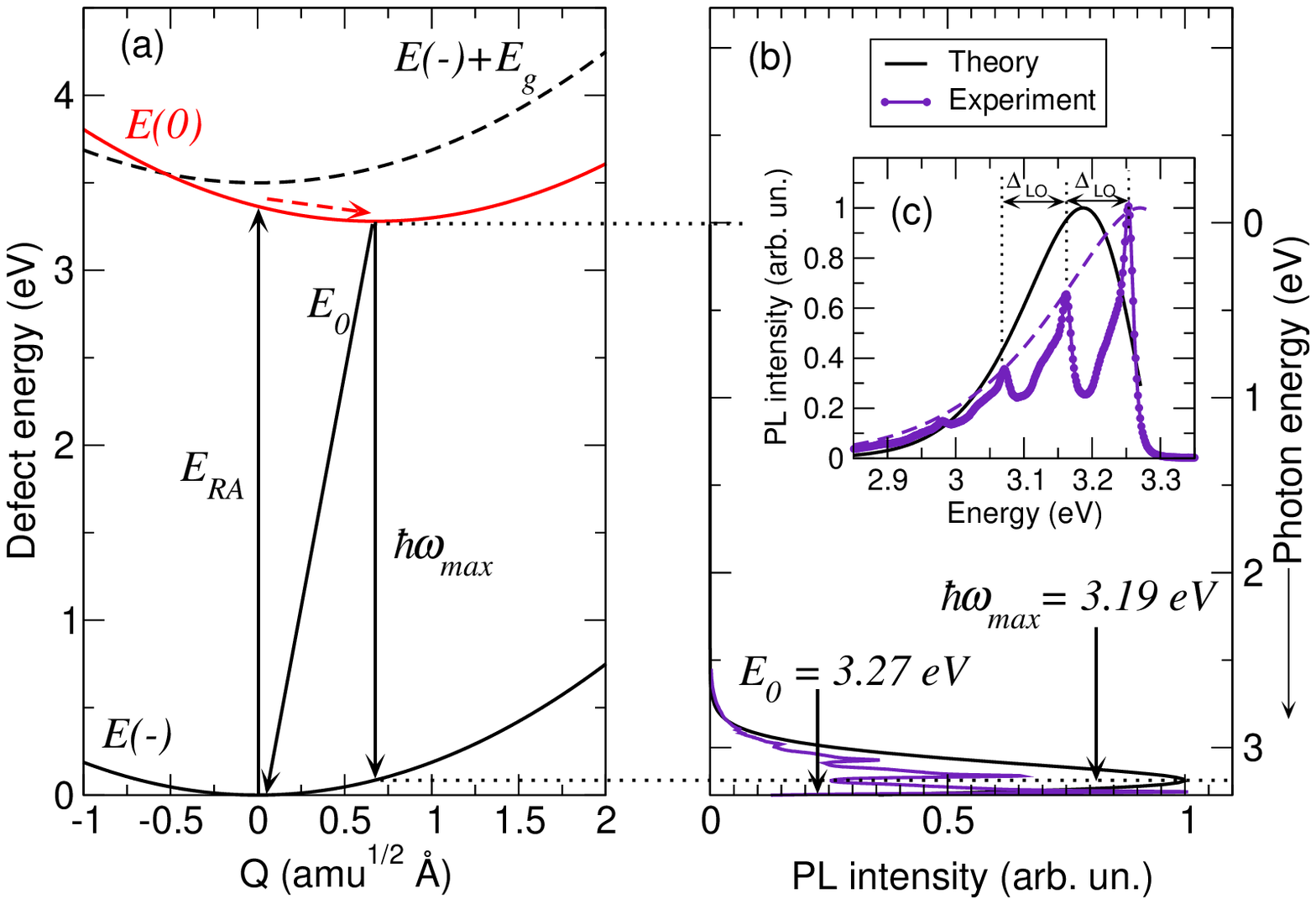}
\caption{\label{fig:Mg-shallow} Calculated configuration coordinate diagram for the shallow state of Mg\textsubscript{Ga} acceptor (a); corresponding HSE predicted PL band shape and UVL\textsubscript{Mg} band (b). The inset (c) shows experimental PL spectrum of UVL\textsubscript{Mg} band (filled symbols), PL band shape calculated from HSE parameters and Eq. (\ref{Eq:I-PL}) (solid line), and obtained from the fit of experimental data into Eq. (\ref{Eq:I-PL}) (dashed line). The fit (dashed line) is obtained using experimental $E_0=3.28$ eV and a fitting parameter $S_e=$ 0.4. Experimental phonon replicas are separated by the $\Delta_{LO}$=92 meV, the energy of the LO phonon mode in bulk GaN.}
\end{figure}

Figure \ref{fig:Mg-shallow} shows the HSE calculated configuration coordinate diagram for the shallow $C_{2v}$ state of Mg\textsubscript{Ga} acceptor and the corresponding calculated PL band shape. It is compared with the experimental UVL\textsubscript{Mg} band commonly observed in Mg-doped GaN samples. Upon photoexcitation, the shallow state of Mg\textsubscript{Ga} acceptor efficiently captures a hole without a barrier (dashed arrow in Fig. \ref{fig:Mg-shallow}(a)). The following radiative transitions are predicted to form a sharp (Fig. \ref{fig:Mg-shallow}(b)) ultraviolet PL band with a ZPL of 3.27 eV, very close to the experimental value of 3.28 eV. The inset (Fig. \ref{fig:Mg-shallow}(c)) shows the details of the UVL\textsubscript{Mg} band shape. As in the case of the shallow state of UVL\textsubscript{Be3}, phonon replicas are separated by $\Delta_{LO}=$92 meV, the energy of the bulk GaN LO phonon mode. 
Using the experimental value of $S_e$=0.4, a fit into the experiment using Eq. (\ref{Eq:I-PL}) is obtained (dashed line in Fig. \ref{fig:Mg-shallow}(c)). In the case of the shallow defect state PL, such a fit describes an envelope of the PL spectrum. 

In contrast, a 1-D configuration diagram based on HSE total energies predicts the ground state vibrational energy (i.e., phonon replica separation) $\hbar \Omega_g=$40 meV, with a Huang-Rhys factor $S_g$=2. The calculated Franck-Condon shift, in this case, is 0.085 eV, which is about a factor of two larger than that of the shallow state of Be\textsubscript{Ga}. 
The calculated vibrational energy of the excited state is very similar to that of the ground state $\hbar \Omega_e=$40 meV with $S_e$=2, due to the similarity of the adiabatic potentials. 
In the experiment, there is almost zero lattice relaxation energy following the optical transition. This, in addition to the difference in vibronic parameters, leads to a slightly redshifted theoretical PL band shape compared to the experiment (Fig. \ref{fig:Mg-shallow}(c)). 
Again, as in the case of Be\textsubscript{Ga} acceptor, defect energetics obtained from HSE are in good agreement with the experiment, while the PL band shapes, based on 1-D effective mode in the direction between R(0) and R(1-) lattice configurations, show some discrepancies. 

\subsection{\label{sec:Zn-level2} Zn\textsubscript{Ga}}

Zn\textsubscript{Ga} acceptor is known to produce the blue PL band in GaN, labeled in the literature as the BL1 or BL\textsubscript{Zn} band \cite{Demchenko_Zn_Ga, Resh_SPIE}. 
For Zn\textsubscript{Ga} acceptor, HSE results yield both $C_{3v}$ and $C_{2v}$ states that are deep. They are predicted by HSE to produce somewhat different PL bands. Below we compare these predicted PL bands to the experimental BL\textsubscript{Zn} band in Zn-doped GaN to determine which of the two states, if not both, is more likely to be the origin of the BL\textsubscript{Zn}.

\subsubsection{Deep $C_{3v}$ state of Zn\textsubscript{Ga}}

\begin{figure}
\hspace{-0.28in}
\vspace{-0.1in}
\includegraphics[scale=0.32]{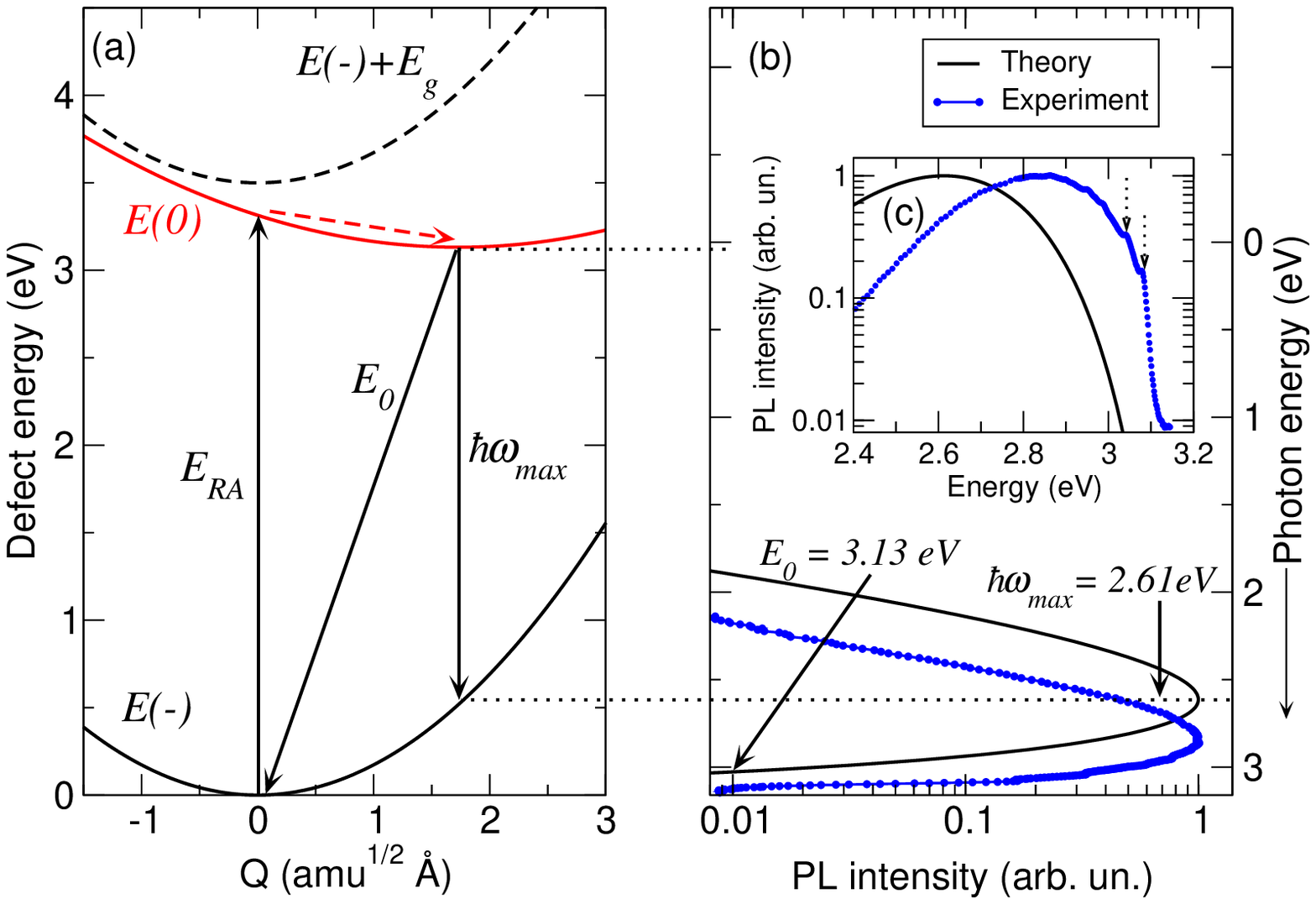}
\caption{\label{fig:Zn} Calculated configuration coordinate diagram for the deep $C_{3v}$ state of Zn\textsubscript{Ga} acceptor (a); the corresponding HSE predicted PL band shape and BL\textsubscript{Zn} band (b). The inset (c) shows the details of the BL\textsubscript{Zn} band shape measured in Zn-doped GaN (filled symbols), band shape calculated from HSE energies and Eq. (\ref{Eq:I-PL}) (solid line). Experimental ZPL and the first phonon replica are separated by 36 meV, as indicated by dashed arrows.}
\end{figure}

Figure \ref{fig:Zn} shows the HSE calculated configuration coordinate diagram for the $C_{3v}$ state of Zn\textsubscript{Ga} acceptor, the corresponding calculated PL band shape, and experimental BL\textsubscript{Zn} band. Resonant absorption by this state is calculated at 3.31 eV. 
Similar to the case of Mg\textsubscript{Ga} acceptor, the adiabatic potentials of the neutral $E(0)$ and negative plus and electron-hole pair $E(-)+E_g$ charge states do not intersect. However, even a small 0.05 eV difference in the HSE calculated resonant absorption $E_{RA}$ would result in the $E(0)$ adiabatic potential that would intersect with the $E(-)+E_g$ potential with a very low (0.07 eV) barrier for the nonradiative hole capture. Therefore the nonradiative hole capture path might be open. 

Following a nonradiative hole capture (dashed arrow in Fig. \ref{fig:Zn}(a)) the PL band with a maximum at 2.61 eV and ZPL at 3.13 eV ($0/-$ transition level at 0.37 eV) is predicted. 
The calculated vibronic parameters of the deep $C_{3v}$ state of Zn\textsubscript{Ga} are $\hbar \Omega_g=$38 meV, $S_g$=14, $\hbar \Omega_e=$23 meV, $S_e$=8. Calculated values of transition energies are in some disagreement with the experimentally observed BL\textsubscript{Zn} band in Zn-doped GaN (Fig. \ref{fig:Zn}(b-c)). Experimental ZPL is 3.1 eV \cite{Resh_SPIE}, as shown in Fig. \ref{fig:Zn}(c) as a sharp peak, indicated by the dashed arrow on the high-energy side of the spectrum. The second dashed arrow indicates the first phonon replica separated by 36 meV. PL maximum of the experimental BL\textsubscript{Zn} band is at 2.86 eV, which differs from the HSE calculated value by 0.25 eV. When compared to the results for the $C_{2v}$ state (Sec. III.C.2), this suggests that the PL band from the deep $C_{3v}$ state is not present in experimental spectra. The reasons for this are likely similar to those in the case of the deep state of Mg\textsubscript{Ga} acceptor and are discussed in Sec. IV. 

\subsubsection{Deep $C_{2v}$ state of Zn\textsubscript{Ga}}

\begin{figure}
\hspace{-0.28in}
\vspace{-0.1in}
\includegraphics[scale=0.32]{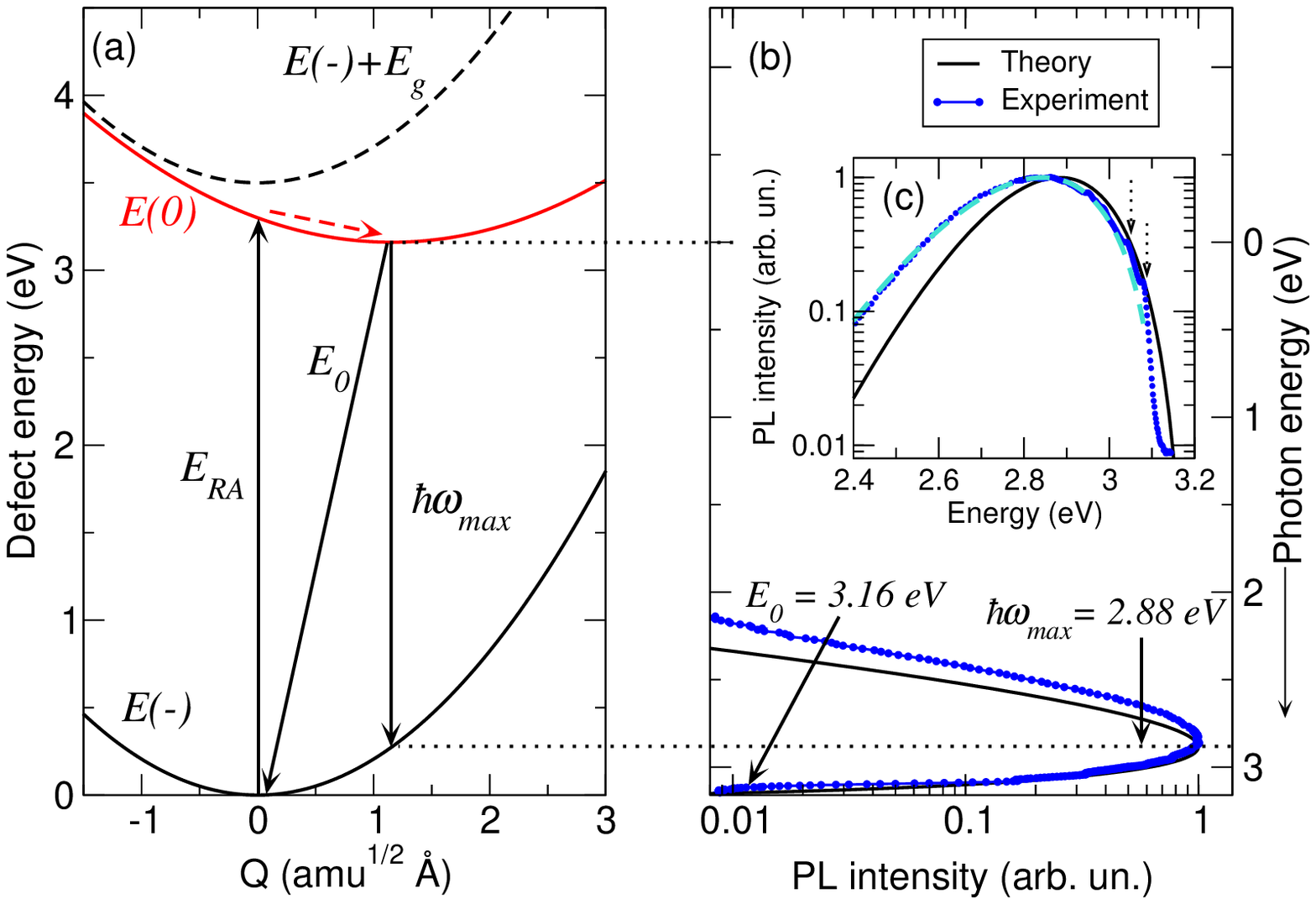}
\caption{\label{fig:Zn-C2v} Calculated configuration coordinate diagram for the $C_{2v}$ state of Zn\textsubscript{Ga} acceptor (a); the corresponding HSE predicted PL band shape and measured BL\textsubscript{Zn} band (b). The inset (c) shows the details of the band shape of BL\textsubscript{Zn} measured in Zn-doped GaN (filled symbols), calculated PL band shape from HSE energies and Eq. (\ref{Eq:I-PL}) (solid line), and the band shape obtained from the fit of experimental data into Eq. (\ref{Eq:I-PL}) (dashed line) with experimental transition energies and the fitting parameter $S_e$=3.2. Experimental ZPL and the first phonon replica are separated by 36 meV, indicated by dashed arrows.}
\end{figure}

Figure \ref{fig:Zn-C2v}(a) shows the HSE calculated configuration coordinate diagram for the $C_{2v}$ state of Zn\textsubscript{Ga} acceptor and the corresponding calculated PL band shape and measured BL\textsubscript{Zn} band (Fig. \ref{fig:Zn-C2v}(b)). The inset shows the details of the experimental BL\textsubscript{Zn} spectrum (filled symbols), compared with the band shape predicted from the HSE computed energies and Eq. (\ref{Eq:I-PL}) (solid line), and the band shape obtained by the fit of Eq. (\ref{Eq:I-PL}) into the experimental data (dashed line, which nearly coincides with experimental data in Fig. \ref{fig:Zn-C2v}(c)). 
Resonant absorption by this state of Zn\textsubscript{Ga} is calculated at 3.30 eV. The adiabatic potentials of the neutral $E(0)$ and negative plus an electron-hole pair $E(-)+E_g$ charge states also do not intersect in this case. However, as in the case of the deep $C_{3v}$ state of Zn\textsubscript{Ga} acceptor, a typical small error in HSE computed absorption transition $E_{RA}$ would result in the $E(0)$ adiabatic potential that exhibits no barrier for the nonradiative hole capture. 
Following a nonradiative hole capture (dashed arrow in Fig. \ref{fig:Zn-C2v}(a)) a PL band with a maximum at 2.88 eV and ZPL at 3.16 eV ($0/-$ transition level is 0.34 eV) is predicted. In this case, calculated transition energies are in good agreement with the experimentally observed BL\textsubscript{Zn} band in Zn-doped GaN (Fig. \ref{fig:Zn-C2v}(b-c)), ZPL is 3.10 eV and PL maximum at 2.86 eV \cite{Resh_SPIE}. 
The predicted vibrational parameters of the $C_{2v}$ state of Zn\textsubscript{Ga} are $\hbar \Omega_g=$42 meV, $S_g$=6.7, $\hbar \Omega_e=$30 meV, $S_e$=4.7. 
Using these values for the theoretical prediction of the PL band shape produces a slightly narrower PL band compared to the experiment (Fig. \ref{fig:Zn-C2v}(c)). 
The value of Huang-Rhys factor $S_e$=3.2 reproduces the experimental band shape (dashed line in Fig. \ref{fig:Zn-C2v}(c)).
The energy separation of the first phonon replica of 36 meV is the vibrational energy of the ground state $\hbar \Omega_g$. PL measurements also reveal a second set of phonon replicas with $\hbar \Omega_g=$ 92 meV, i.e. bulk GaN LO phonon mode \cite{Reshchikov-Morkoc}. 
Temperature dependence of PL band width yields the value of $\hbar \Omega_e=$43 meV \cite{Reshchikov-Morkoc}. This suggests that the vibrational properties that determine the PL band shape are defined by a combination of the bulk LO phonon mode and a local vibrational mode.  
Overall these results indicate that the BL\textsubscript{Zn} band originates from the deep $C_{2v}$ state of Zn\textsubscript{Ga}.

\subsection{\label{sec:Ca-level2} Ca\textsubscript{Ga}}

Ca\textsubscript{Ga} acceptor has been suggested as the origin of the green luminescence band labeled GL\textsubscript{Ca} in GaN samples implanted or unintentionally doped with Ca \cite{Resh_Ca_Ga,Resh_SPIE}. Our previous calculations of optical transitions via the deep $C_{3v}$ state of Ca\textsubscript{Ga} acceptor \cite{Resh_Ca_Ga} showed a discrepancy of 0.25 eV from the experiment. In addition, we were unable to locate a stable $C_{2v}$ state for this defect. Here we revisit these results with a revised parametrization of HSE, as described in Sec. II.B. We also do not find a stable $C_{2v}$ state of Ca\textsubscript{Ga} using this revised approach.

\begin{figure}
\hspace{-0.28in}
\vspace{-0.1in}
\includegraphics[scale=0.32]{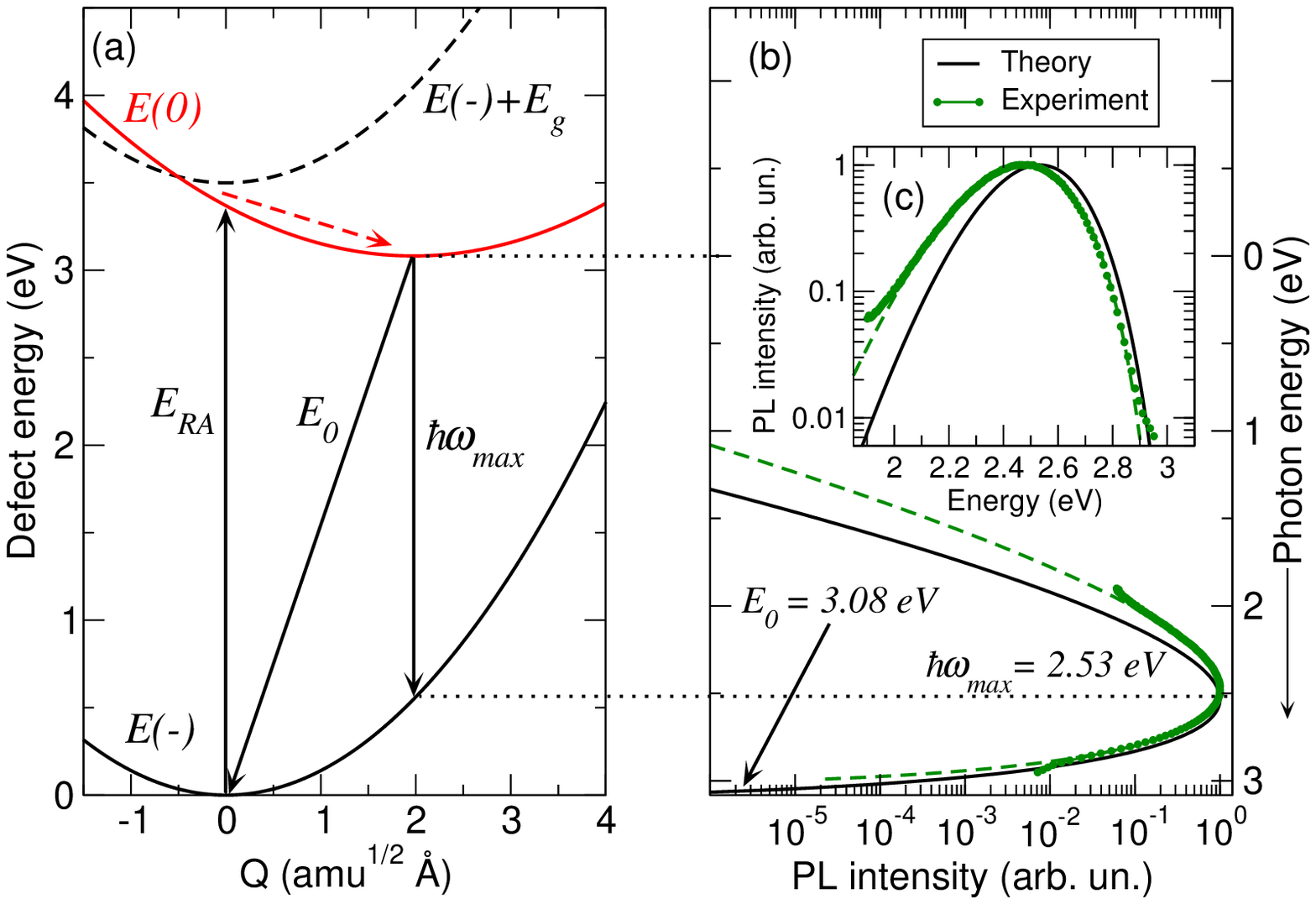}
\caption{\label{fig:Ca} Calculated configuration coordinate diagram for the $C_{3v}$ state of Ca\textsubscript{Ga} acceptor (a); corresponding HSE predicted PL band shape (solid line), experimental GL\textsubscript{Ca} band shape (filled symbols), and the fit of Eq. (\ref{Eq:I-PL}) into experiment (b). The inset (c) shows the details of the spectrum of GL\textsubscript{Ca} band measured in Ca-implanted GaN (filled symbols), the band shape calculated from HSE total energies using Eq. (\ref{Eq:I-PL}) (solid line) PL, and the fit of experimental data into Eq. (\ref{Eq:I-PL}) (dashed line) with experimental transition energies and the fitting parameter $S_e=$ 8.5.}
\end{figure}

Figure \ref{fig:Ca}(a) shows the HSE calculated configuration coordinate diagram for the $C_{3v}$ state of Ca\textsubscript{Ga} acceptor and the corresponding calculated PL band shape along with experimental GL\textsubscript{Ca} and the fit of Eq. (\ref{Eq:I-PL}) into experiment(Fig. \ref{fig:Ca}(b)).  
The inset shows the details of the experimental GL\textsubscript{Ca} spectrum (filled symbols), compared with the band shape predicted from the HSE computed total energies using Eq. (\ref{Eq:I-PL}) (solid line), and the band shape obtained by the fit of Eq. \ref{Eq:I-PL} into the experimental data (dashed line in Fig. \ref{fig:Ca}(c)). 
Resonant absorption by this state is calculated at 3.37 eV. The adiabatic potentials of the neutral and negative plus and electron-hole pair charge states suggest an efficient nonradiative hole capture (dashed arrow in Fig. \ref{fig:Ca}(a)) without a barrier. 
The PL band with a maximum at 2.53 eV and ZPL at 3.08 eV ($0/-$ transition level is 0.42 eV) is predicted. The calculated transition energies are in good agreement with the experimentally observed GL\textsubscript{Ca} band in Ca-doped GaN (Fig. \ref{fig:Ca}(b-c)), ZPL is 3.0 eV and PL maximum at 2.5 eV \cite{Resh_Ca_Ga, Resh_SPIE}. 
The predicted vibrational parameters of the $C_{3v}$ state of Ca\textsubscript{Ga} are $\hbar \Omega_g=$34 meV, $S_g$=16, $\hbar \Omega_e=$25 meV, $S_e$=11.6. Using these values for the theoretical prediction of the PL band shape produces a slightly narrower PL band compared to the experiment (Fig. \ref{fig:Ca}(c)). 
Measurements of temperature dependence of the PL band width reveals $\hbar \Omega_e=$41 meV \cite{Resh_Ca_Ga}. 
The experimental PL band shape is reproduced with $S_e$=8.5 (dashed line in Fig. \ref{fig:Ca}(c)). In contrast to the defects discussed above, the PL spectrum from the Ca\textsubscript{Ga} acceptor does not show fine structure (i.e. resolved ZPL and phonon replicas) possibly because of the ZPL and phonon replicas' broadening due to implantation damage. 
 
\subsection{\label{sec:Cd-level2} Cd\textsubscript{Ga}}

\begin{figure}
\hspace{-0.28in}
\vspace{-0.1in}
\includegraphics[scale=0.32]{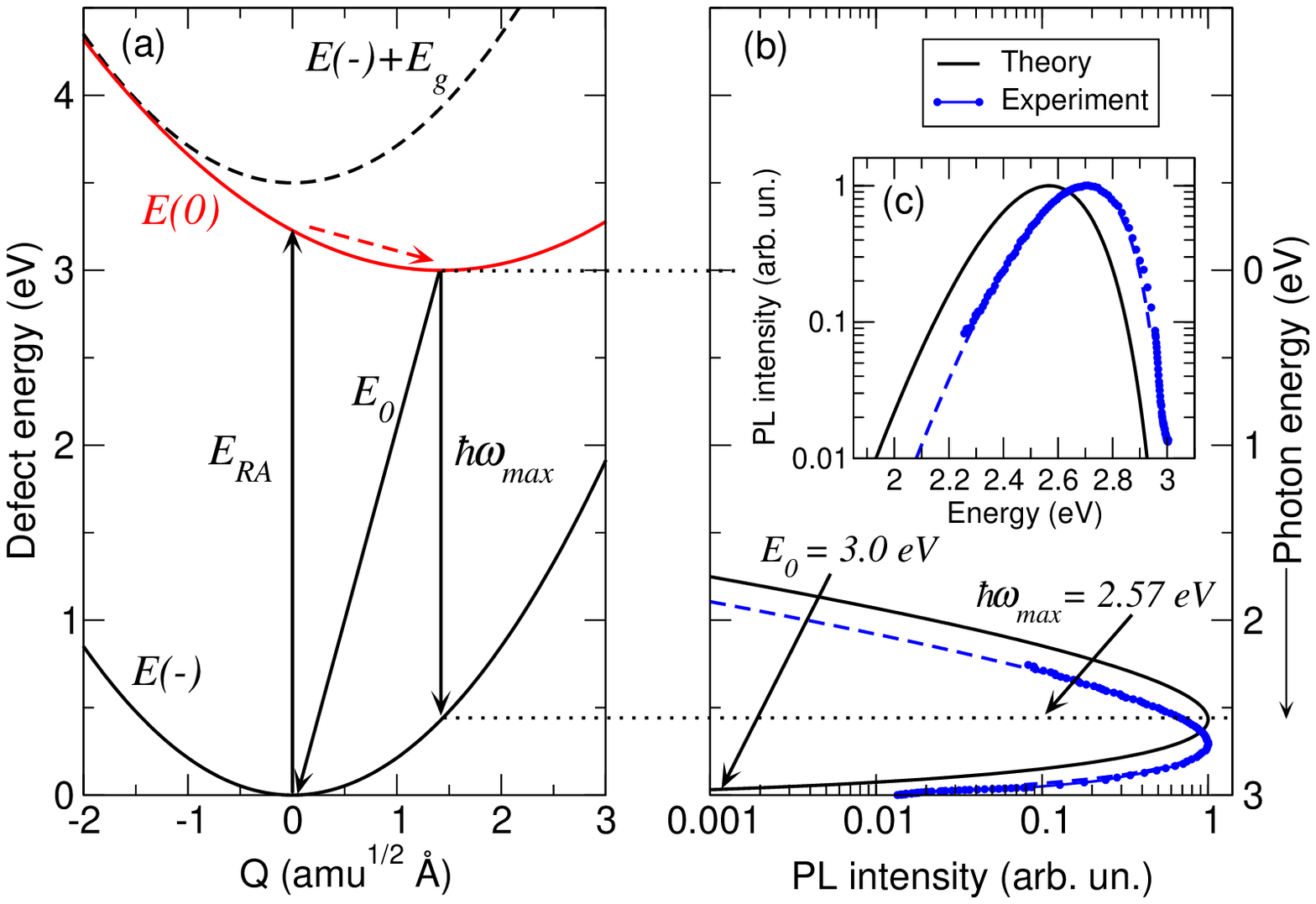}
\caption{\label{fig:Cd} Calculated configuration coordinate diagram for the $C_{3v}$ state of Cd\textsubscript{Ga} acceptor (a); corresponding HSE predicted PL band shape, experimental BL\textsubscript{Cd} band (filled symbols), and the fit of the experiment into Eq. (\ref{Eq:I-PL}) (dashed line) (b). The inset (c) shows the details of the band shape of BL\textsubscript{Cd} band measured in Cd-doped GaN (filled symbols), PL band shape calculated from HSE energies and Eq. (\ref{Eq:I-PL}) (solid line), and the band shape obtained from the fit of experimental data into Eq. (\ref{Eq:I-PL}) (dashed line) with experimental transition energies and fitting parameter $S_e=$ 4.0.}
\end{figure}

HSE calculations of Cd\textsubscript{Ga} acceptor also do not yield a stable $C_{2v}$ state for this defect. 
Figure \ref{fig:Cd}(a) shows the HSE calculated configuration coordinate diagram for the $C_{3v}$ state of Cd\textsubscript{Ga} acceptor and the corresponding calculated PL band shape (solid line), compared with the experimental BL\textsubscript{Cd} band (filled symbols), and the fit of Eq. (\ref{Eq:I-PL}) into the experiment (dashed line) (Fig. \ref{fig:Cd}(b)). 
The inset shows the details of the experimental BL\textsubscript{Cd} spectrum (filled symbols), compared with the band shape predicted from Eq. (\ref{Eq:I-PL}) using the HSE computed energies (solid line), and the band shape obtained by the fit of Eq. (\ref{Eq:I-PL}) into the experimental data (Fig. \ref{fig:Cd}(b,c)). 

Resonant absorption by this state is calculated at 3.23 eV. The adiabatic potentials of the neutral $E(0)$ and negative plus and electron-hole pair $E(-)+E_g$ charge states show a barrier for the nonradiative hole capture (dashed arrow in Fig. \ref{fig:Cd}(a)) of about 0.3 eV. However, small inaccuracies in calculated absorption energy or the value of the ZPL can lead to slightly different adiabatic potentials with significant variations in this barrier. 
Since HSE calculations can produce deviations from the experiment of about 0.1 eV for PL maximum (Fig. \ref{fig:Cd}(c)), the existence of this barrier is inconclusive. 
HSE predicts the PL band with a maximum at 2.57 eV and ZPL at 3.0 eV. The calculated transition energies are in reasonable agreement with the experimentally observed BL\textsubscript{Cd} band in Cd-doped GaN (Fig. \ref{fig:Cd}(b-c)), where measured ZPL is 2.95 eV and PL maximum is at 2.7 eV \cite{Resh_Cd-Hg, Bergman_Zn-Cd}. 
The calculated vibronic parameters of the $C_{3v}$ state of Cd\textsubscript{Ga} are $\hbar \Omega_g=$42 meV, $S_g$=10, $\hbar \Omega_e=$31 meV, $S_e$=7.5. Using these values for the theoretical prediction of the PL band shape produces a PL band with a similar shape to that observed in the experiment (Fig. \ref{fig:Cd}(c)), albeit redshifted by about 0.1 eV. 
The experimental PL band shape is best reproduced with $S_e$=4.0 (dashed line in Fig. \ref{fig:Cd}(c)). Temperature dependence of the PL band width yields $\hbar \Omega_e=$53 meV \cite{Resh_Cd-Hg}. 
In this case, as for several acceptors discussed above, HSE calculated transition energies are reproduced reasonably well, while the calculated vibrational parameters show some discrepancy. 

\subsection{\label{sec:Hg-level2} Hg\textsubscript{Ga}}

\begin{figure}
\hspace{-0.28in}
\vspace{-0.1in}
\includegraphics[scale=0.32]{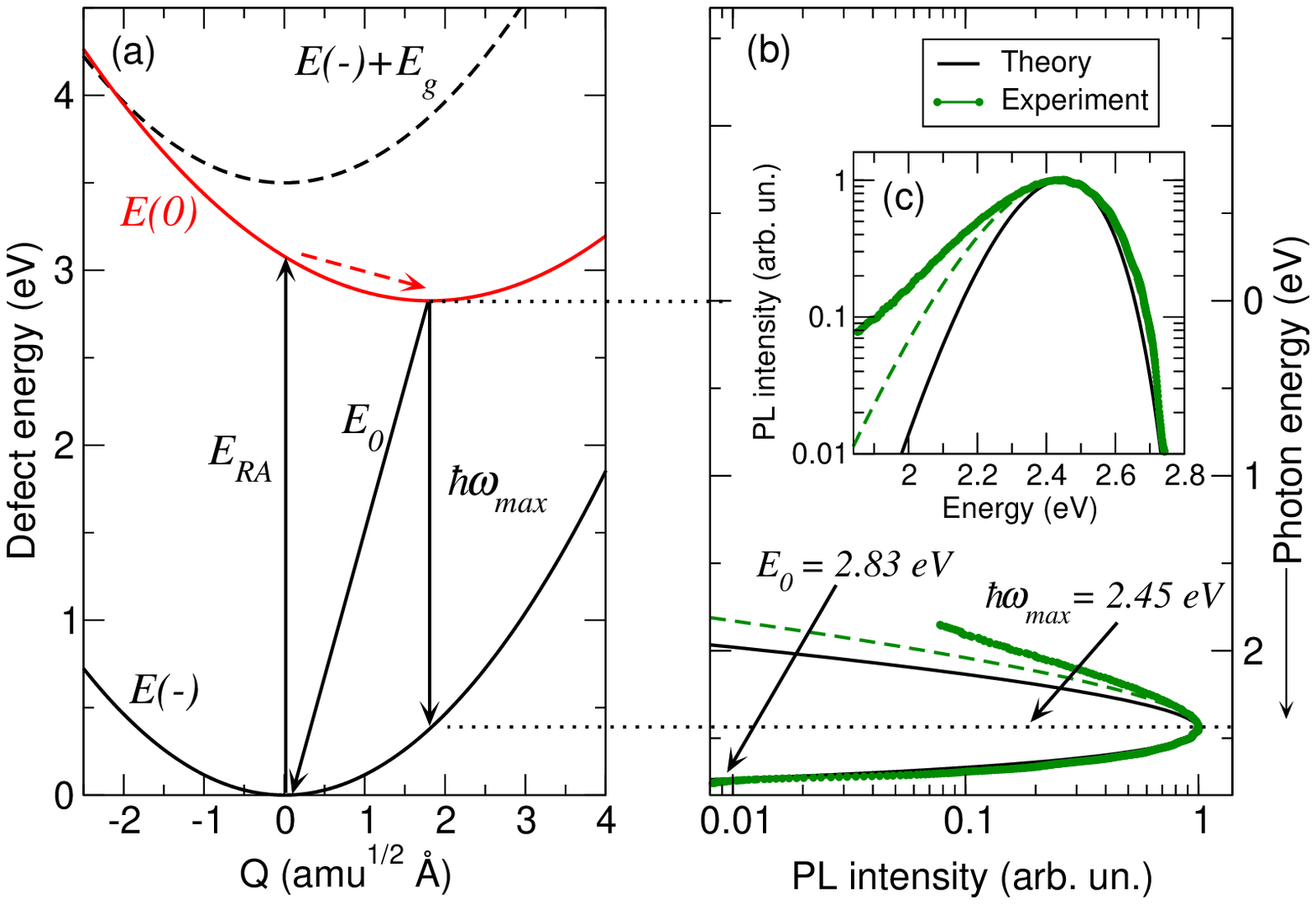}
\caption{\label{fig:Hg} Calculated configuration coordinate diagram for the $C_{3v}$ state of Hg\textsubscript{Ga} acceptor (a); corresponding HSE predicted PL band shape, experimental GL\textsubscript{Hg} band (filled symbols), and the fit of the experiment into Eq. (\ref{Eq:I-PL}) (dashed line) (b). The inset (c) shows the details of the band shape of GL\textsubscript{Hg} band measured in Hg-doped GaN (filled symbols), PL band shape calculated from HSE energies and Eq. (\ref{Eq:I-PL}) (solid line), and the band shape obtained from the fit of experimental data into Eq. (\ref{Eq:I-PL}) (dashed line) with experimental transition energies and fitting parameter $S_e=$ 5.0.}
\end{figure}

HSE calculations of Hg\textsubscript{Ga} acceptor also do not produce a stable $C_{2v}$ state. 
Figure \ref{fig:Hg}(a) shows the HSE calculated configuration coordinate diagram for the $C_{3v}$ state of Hg\textsubscript{Ga} acceptor and the corresponding calculated PL band shape (Fig. \ref{fig:Hg}(b)). 
The inset shows the details of the experimental Hg\textsubscript{Cd} spectrum (filled symbols), compared with the band shape predicted from Eq. (\ref{Eq:I-PL}) using the HSE computed energies (solid line), and the band shape obtained by the fit of Eq. (\ref{Eq:I-PL}) into the experimental data (Fig. \ref{fig:Hg}(c)). 

Resonant absorption by this state is calculated at 3.08 eV. The adiabatic potentials of the neutral $E(0)$ and negative plus and electron-hole pair $E(-)+E_g$ charge states show a barrier for the nonradiative hole capture (dashed arrow in Fig. \ref{fig:Hg}(a)) of about 0.5 eV, suggesting temperature dependence of the hole capture coefficient. 
However, as in the above-discussed cases of Zn\textsubscript{Ga} and Cd\textsubscript{Ga} acceptors, the value of this barrier is sensitive to small variations in the HSE calculated transition energies. In addition, this barrier might be different if a different local or pseudo-local mode is dominating the nonradiative transitions.  
HSE predicts the PL band with a maximum at 2.45 eV and ZPL at 2.83 eV. 
The calculated transition energies are in good agreement with the experimentally observed GL\textsubscript{Hg} band in Hg-doped GaN (Fig. \ref{fig:Hg}(b-c)), where measured ZPL is 2.73 eV and PL maximum is at 2.44 eV \cite{Resh_Cd-Hg}. The calculated vibronic parameters of the $C_{3v}$ state of Hg\textsubscript{Ga} are $\hbar \Omega_g=$31 meV, $S_g$=12, $\hbar \Omega_e=$25 meV, $S_e$=10. Using these values for the theoretical prediction of the PL band shape once again produces a narrower PL band compared to the experiment (Fig. \ref{fig:Hg}(c)). 
The experimental PL band shape is well reproduced with $S_e$=5 (dashed line in Fig. \ref{fig:Hg}(b,c)). Temperature measurements of the PL band width yield $\hbar \Omega_e=$53 meV, similar to the case of Cd\textsubscript{Ga} \cite{Resh_Cd-Hg}.
The low-energy side of the PL spectrum shows some discrepancy between the fit of experimental data into Eq. (\ref{Eq:I-PL}) and the experimental spectrum. This is an example where a 1-D configuration coordinate diagram fails to fully reproduce the PL band shape. 
As for most acceptors discussed above, HSE calculated transition energies are well reproduced, while the PL band shape of the Hg\textsubscript{Ga} shows some disagreement with the experiment.

\section{Dual nature of acceptors in PL experiments}
Among acceptors considered in this work, Be\textsubscript{Ga}, Mg\textsubscript{Ga}, and Zn\textsubscript{Ga} are predicted to have both  $C_{2v}$ and $C_{3v}$ states. For Be\textsubscript{Ga} and Mg\textsubscript{Ga} acceptors, $C_{2v}$ states are shallow, with PL bands typical for shallow acceptors, i.e. sharp PL band shapes with a series of well-defined phonon replicas shown in Figs. (\ref{fig:Be-shallow}, \ref{fig:Mg-shallow}). For Zn\textsubscript{Ga} acceptor, both $C_{2v}$ and $C_{3v}$ states are deep and localized within the supercell used here. However, only for Be\textsubscript{Ga} both states produce a measurable PL. The deep $C_{3v}$ states of Mg\textsubscript{Ga} and Zn\textsubscript{Ga} acceptors are not observed in PL experiments. 

In order to identify trends in the behavior of cation site acceptors in GaN, Table \ref{Table:Atom-radii} shows the atomic radii $R_a$ for acceptor impurity atoms and corresponding acceptor transition levels calculated using the Koopmans tuned HSE in this work. The atomic radius $R_a$ serves as a rough estimate of the core atom size in the calculation. It is obtained from the PAW pseudopotential fitting, where the logarithmic derivatives of the atomic all-electron wavefunction are matched to the PAW pseudo-wavefunction at the distance $R_a$ from the atom center. The comparison of the results shown in Table \ref{Table:Atom-radii} reveals two chemical trends in the formation of Ga-site acceptors in GaN. 
One is that larger atoms tend to create deeper acceptor levels in the bandgap. 
Another is that the atoms roughly comparable in size to the Ga atom they substitute for, do not form stable $C_{2v}$ acceptor states. Only deep polaronic $C_{3v}$ and pseudo-$C_{3v}$ acceptor states are formed by Ca\textsubscript{Ga}, Cd\textsubscript{Ga}, and Hg\textsubscript{Ga} acceptors. 
This is also revealed in the experiment by wide PL bands originating from these acceptors. 
An intermediate case is Zn\textsubscript{Ga} acceptor, where atomic radius only slightly smaller than that of the Ga atom, forms $C_{3v}$, pseudo-$C_{3v}$, and $C_{2v}$ deep acceptor states. 
In contrast, atoms with $R_a$ smaller than that of Ga, i.e. Be and Mg form stable (or metastable in case of Be\textsubscript{Ga}) shallow $C_{2v}$ states, that are revealed as narrow PL bands with sharp ZPLs and resolved phonon replicas in the experiment.  

\begin{table}[t]
\caption{\label{Table:Atom-radii}%
Atomic radii $R_a$(\AA) obtained from PAW pseudopotential fitting for acceptor impurity atoms substituting for Ga, and HSE calculated 0/- acceptor transition levels (eV) above the VBM for $C_{2v}$ and $C_{3v}$ acceptor states. The atomic radius for the Ga atom is given for comparison.}
\begin{ruledtabular}
\begin{tabular}{c c c c}
Atom &
$R_a$ & 0/- ($C_{3v}$)& 0/- ($C_{2v}$)\\
\colrule
Be & 1.005 & 0.34 & 0.18 \\
Mg & 1.058 & 0.21 & 0.23 \\
Zn & 1.270 & 0.37 & 0.34 \\
Ca & 1.482 & 0.42 & N/A \\
Cd & 1.482 & 0.50 & N/A \\
Hg & 1.376 & 0.67 & N/A \\
Ga & 1.402 & & \\
\end{tabular}
\end{ruledtabular}
\end{table}

Observation (or existence) of the dual nature of acceptors has been a subject of debate for over a decade. While both deep polaronic and shallow states of acceptors were predicted theoretically in the past \cite{Lany-Zunger_dual-nature, Demchenko_Be_PRL, Demchenko_Mg, Raebiger_Mg}, the experimental observation of the two states existing for the same acceptor was demonstrated only recently in Ref. \cite{Resh_dual-nature_Be} for Be\textsubscript{Ga} acceptor. At low temperatures only the deep $C_{3v}$ ground state of Be\textsubscript{Ga} produces the PL. 
At $T>140$ K a shallow $C_{2v}$ state opens as the radiative recombination channel, and the UVL\textsubscript{Be3} emerges. 
The UVL\textsubscript{Be3} band with the main peak at 3.26 eV should not be confused with the UVL\textsubscript{Be} band with a peak at 3.38 eV observed in GaN samples with high concentrations of Be. The latter originates from the shallowest acceptor in GaN known to date \cite{Demchenko_Be_PRL, Demchenko_Be_APL}, which remains unidentified. 
(Note, in the two references above it was assumed that UVL\textsubscript{Be} originated from the shallow state of Be\textsubscript{Ga} acceptor, which is disputed by the latest experiments \cite{Resh_dual-nature_Be}.)

The current experimental picture is that both $C_{3v}$ and $C_{2v}$ states are observed as PL bands for Be\textsubscript{Ga} acceptor, while only the $C_{2v}$ state is observed for Mg\textsubscript{Ga} and Zn\textsubscript{Ga}.
In Ref. \cite{Demchenko_Be_PRL} we argued that the deep polaronic states are not revealed in PL spectra due to significantly lower hole capture coefficients for these states compared to the shallow states. Later in Ref. \cite{Resh_dual-nature_Be} we showed that the energy differences and potential energy barriers between the states competing for hole capture play a major role in which state dominates the PL spectrum at different temperatures. 

\begin{figure}
%\includegraphics[width=3.4in,height=3.4in]{Barriers.eps}
%\hspace{-0.28in}
\vspace{0.1in}
\includegraphics[scale=0.31]{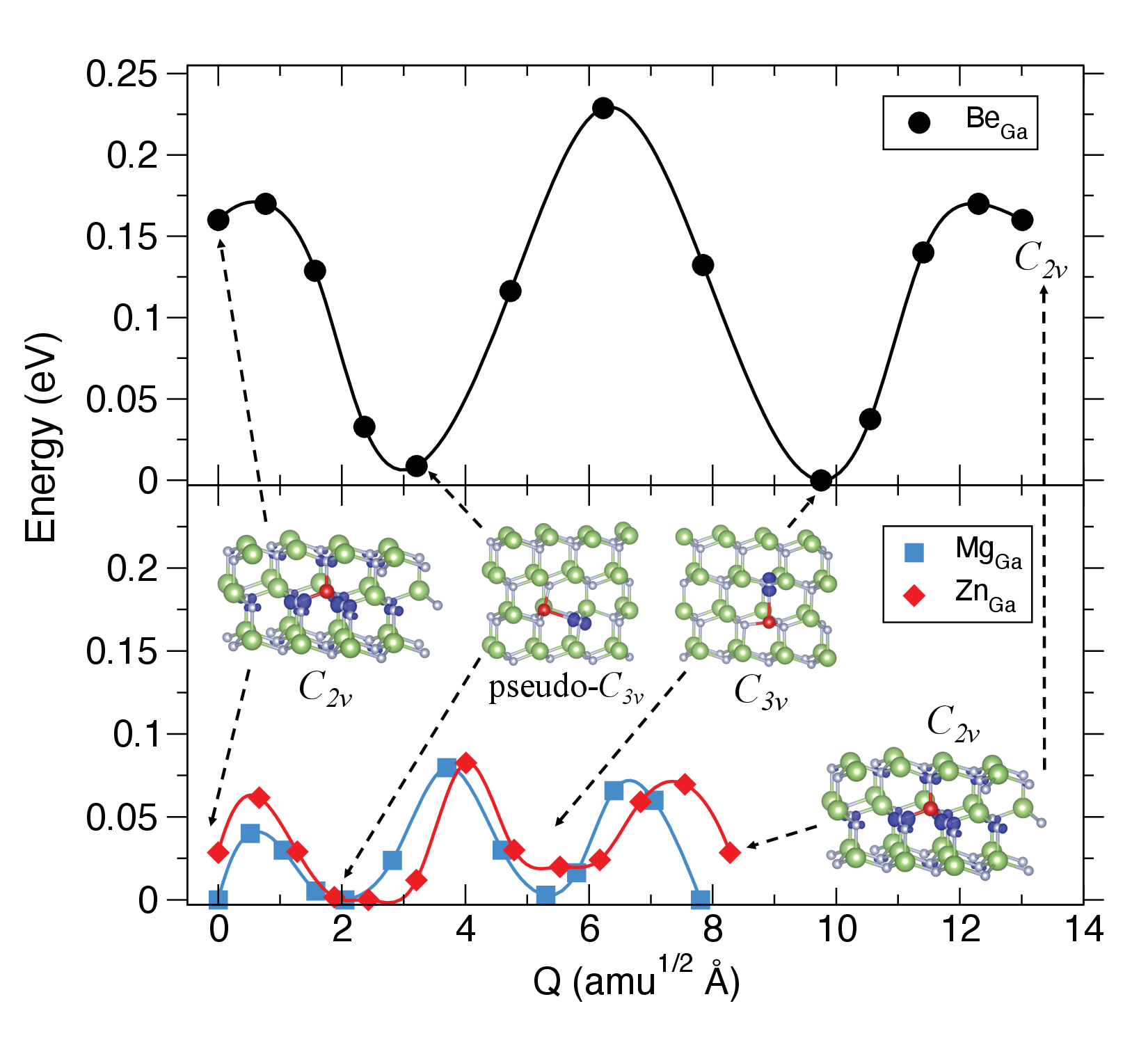}
\caption{\label{fig:Barriers} Total energies of the three configurations of neutral Be\textsubscript{Ga}, Mg\textsubscript{Ga}, and Zn\textsubscript{Ga} acceptors, i.e. $C_{2v}$, pseudo-$C_{3v}$, and $C_{3v}$ states of acceptors, and the potential barrier paths between them as a function of the configuration coordinate $Q$. The origin $Q=0$ corresponds to the $C_{2v}$ state for each acceptor, the first minima are the pseudo-$C_{3v}$ states of acceptors, the second minima are the $C_{3v}$ states, and the last points are the $C_{2v}$.}
\end{figure}

Here, we use Koopmans tuned HSE to calculate the energy barriers between the three different states of Be\textsubscript{Ga}, Mg\textsubscript{Ga}, and Zn\textsubscript{Ga} acceptors (shown in Fig. (\ref{fig:WF})). The calculations were performed using the nudged elastic method in 300-atom supercells, with the HSE functional tuned for the deep $C_{3v}$ ground state of each acceptor. Since these functionals do not appropriately fulfill the Koopmans' condition for $C_{2v}$ states, total energy corrections were applied in post-processing to bring the total energy differences between the states in accord with the thermodynamic transition levels calculated by the appropriate HSE functionals. 
Figure \ref{fig:Barriers} shows the total energies between the three configurations of the neutral acceptors as a function of the configuration coordinate $Q$. While the potential landscapes for Mg\textsubscript{Ga} and Zn\textsubscript{Ga} acceptors are very similar, the energy differences and potential barriers between acceptor states for Be\textsubscript{Ga} are significantly larger. 

Total energy differences between $C_{2v}$, (pseudo)-$C_{3v}$, and $C_{3v}$ states for Mg\textsubscript{Ga} and Zn\textsubscript{Ga} are within the error of HSE calculations (i.e. 0.02 eV), indicating bi-stability for both Mg\textsubscript{Ga} and Zn\textsubscript{Ga} acceptors. The potential barriers between these states are between 0.04 and 0.07 eV in Fig. \ref{fig:Barriers}. This suggests that at low temperatures, the state that more efficiently captures the holes (non-radiatively) will dominate the PL spectra since the defect will stay in the state to which the hole is captured. At higher temperatures, since the barriers to and from the deep/shallow states are nearly symmetric, a dynamic equilibrium will be established, populating shallow and deep states equally. Thus, the state that efficiently captures electrons (radiatively) will result in higher-intensity PL. For both Mg\textsubscript{Ga} and Zn\textsubscript{Ga} acceptors this is $C_{2v}$ state, which is observed in the experiment. 

On the other hand, for Be\textsubscript{Ga}, shallow $C_{2v}$ state is 0.16 eV higher in energy than the deep $C_{3v}$ state (from experiment transition level for the deep state is 0.14 eV deeper than that of the shallow state). There is a low ($\sim$0.01 eV) potential barrier from the shallow to the deep state, suggesting an efficient transfer of the captured hole from the shallow to the deep state. Thus, at low temperatures, despite the efficient capture of the photogenerated hole by the shallow state, the deep state will dominate the PL spectrum. At higher temperatures, the thermal emission of holes from the deep to the shallow state would allow for the radiative transitions via the shallow state and increase the intensity of the UVL\textsubscript{Be3} band. 

Temperature dependences of PL intensities associated with shallow and deep states can be computed using the rate equation model derived in Ref. \cite{Resh_dual-nature_Be}. The following expressions can be obtained for the quantum efficiency of the deep-state PL ($\eta_D$) and shallow-state PL ($\eta_S$):
\begin{equation}
\eta_D = \frac{\tau_{D \rightarrow S}}{\tau_{n \rightarrow D}} \frac{B\eta_D(0)+\eta_S(0)}{AB-1}
\label{Eq:Efficiency_D}
\end{equation}
\begin{equation}
\eta_S = \frac{\tau_{D \rightarrow S}}{\tau_{n \rightarrow S}} \frac{\eta_D(0)+A\eta_S(0)}{AB-1}
\label{Eq:Efficiency_S}
\end{equation}
with $A=1+\tau_{D \rightarrow S}(1/\tau_{n \rightarrow D} + 1/\tau_{D,therm})$, and $B=1+\tau_{S \rightarrow D}(1/\tau_{n \rightarrow S} + 1/\tau_{S,therm})$. Here, $\eta_D(0)$ and $\eta_S(0)$ are the PL internal quantum efficiencies (IQE) for the deep and shallow states, respectively, in the limit of low temperatures, $\tau_{n \rightarrow D}$ and $\tau_{n \rightarrow S}$ are PL lifetimes for the deep and shallow states, $\tau_{D \rightarrow S}=\nu^{-1}_D \exp(E_{DS}/kT)$ and $\tau_{S \rightarrow D}=\nu^{-1}_S \exp(E_{SD}/kT)$ are characteristic times of the system transfer from the deep to shallow ($D \rightarrow S$) or shallow to deep ($S \rightarrow D$) states with vibrational frequencies $\nu_D$ and $\nu_S$ over the barriers $E_{DS}$ and $E_{SD}$, respectively. The characteristic times of thermal emission of holes from the deep and shallow states to the valence band $\tau_{D,therm}$ and $\tau_{S,therm}$ have the following expressions: $\tau_{D,therm}=(C_{pD}N_v)^{-1}g\exp(E_D/kT)$ and $\tau_{S,therm}=(C_{pS}N_v)^{-1}g\exp(E_S/kT)$, where $C_{pD}$ and $C_{pS}$ are the hole-capture coefficients for the deep and shallow states, $E_D$ and $E_S$ are the energies of the deep and shallow -/0 transition levels above the VBM, $g$ is the degeneracy factor (assumed for simplicity equal 2), and $N_v$ is the effective density of states in the valence band. In contrast to Ref. \cite{Resh_dual-nature_Be} where the capture of holes by the deep state was ignored, here we account for the hole capture at both deep and shallow states with capture coefficients $C_{pD}$ and $C_{pS}$. 

\begin{figure}[t]
%\includegraphics[width=3.4in,height=3.4in]{Barriers.eps}
%\hspace{-0.28in}
%\vspace{0.1in}
\includegraphics[scale=0.45]{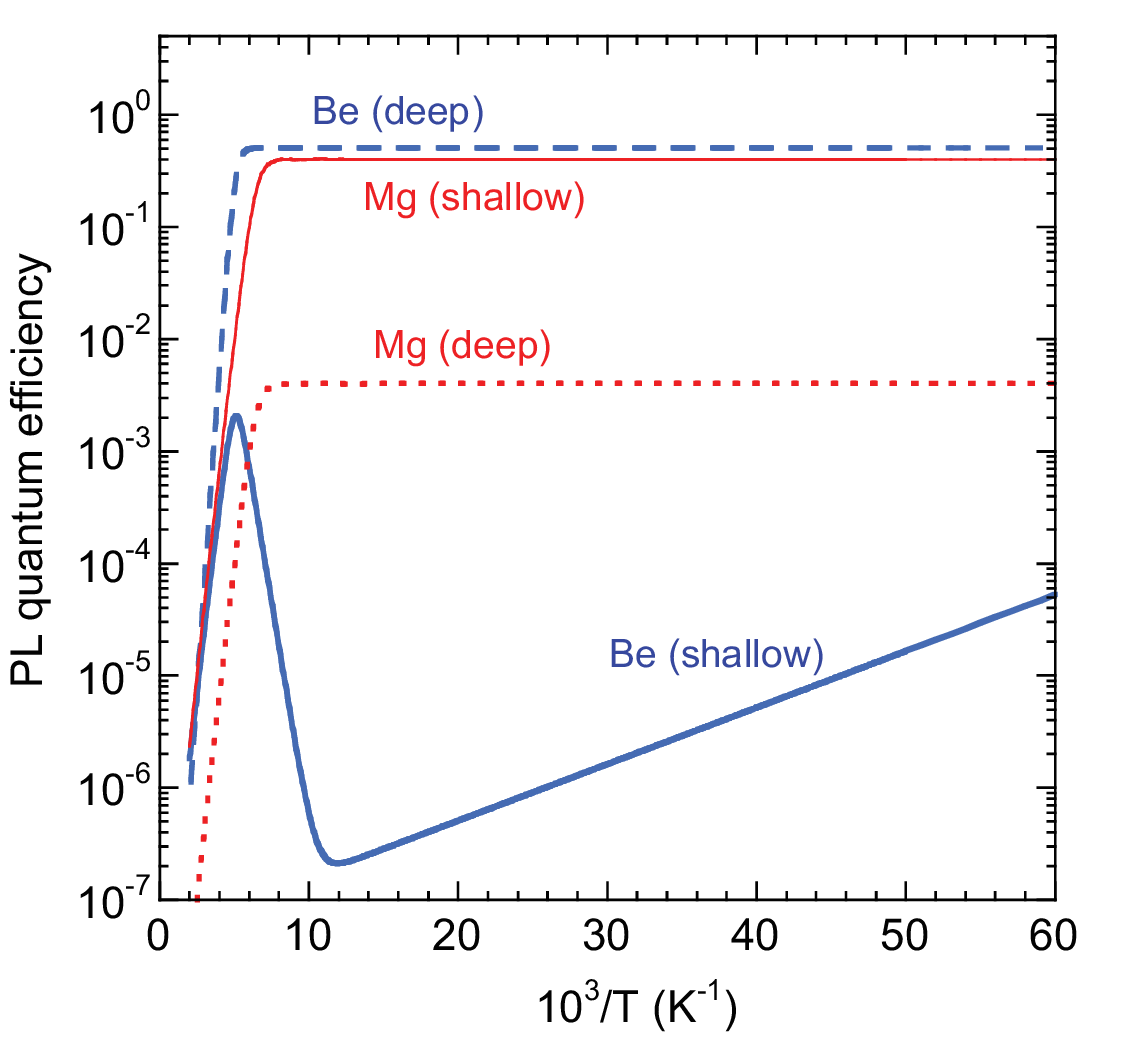}
\caption{\label{fig:Efficiency} Temperature dependences of PL IQE for the shallow and deep states of Be\textsubscript{Ga} and Mg\textsubscript{Ga} acceptors in GaN calculated using Eqs.(\ref{Eq:Efficiency_D},\ref{Eq:Efficiency_S}) with the following parameters: $\tau_{n \rightarrow D}$=10\textsuperscript{-4}s, $\tau_{n \rightarrow S}$=10\textsuperscript{-6}s, $C_{pD}=10^{-8}$cm\textsuperscript{3}/s, $C_{pS}=10^{-6}$cm\textsuperscript{3}/s, $\nu_D=\nu_S$=10\textsuperscript{13}s\textsuperscript{-1} (for both acceptors); 
$E_D$ = 0.34 eV, $E_S$ = 0.18 eV, $E_{DS}$ = 0.17 eV, $E_{SD}$ = 0.01 eV (for Be\textsubscript{Ga});
$E_D$ = $E_S$ = 0.20 eV, $E_{DS}$ = $E_{SD}$ = 0.04 eV (for Mg\textsubscript{Ga}).
}
\end{figure}

Examples of calculated $\eta_D(T)$ and $\eta_S(T)$ dependences for the Be\textsubscript{Ga} and Mg\textsubscript{Ga} acceptors in GaN are shown in Fig. \ref{fig:Efficiency}.  
In the limit of low temperatures, the PL from the shallow state is stronger than that from the deep state by a factor of $C_{pS}/C_{pD}$. 
For the Be\textsubscript{Ga} acceptor with selected parameters PL from the shallow state is expected to be observed at $10^3/T>150 ~K^{-1}$ ($T<7~K$, not shown here). With increasing temperature, the PL intensity from the shallow state decreases as $\exp(E_{SD}/kT)$ for the Be\textsubscript{Ga} acceptor because the potential barrier between the shallow and deep states is low ($E_{SD}$=10 meV) but does not decrease for Mg\textsubscript{Ga} acceptors, for which $E_{SD}$=40 meV. At intermediate temperatures ($10^3/T=10-50 ~K^{-1}$) the transitions between the deep and shallow states of the Mg\textsubscript{Ga} acceptor become significant, but these states are equally populated because their energies are equal. In this temperature region, the PL from the shallow state is stronger than that from the deep state by a factor of $\tau_{n \rightarrow D}/\tau_{n \rightarrow S}$ for the Mg\textsubscript{Ga} acceptor. For the Be\textsubscript{Ga} acceptor, the population of the shallow state with holes decreases up to $T\sim 100~K$ and starts increasing only at $10^3/T<10 ~K^{-1}$ because of a relatively large difference between the deep and shallow levels (calculated $\Delta E=160$ meV, measured $\Delta E=140$ meV). At $10^3/T<7 ~K^{-1}$ for Mg\textsubscript{Ga} and $10^3/T<5 ~K^{-1}$ for Be\textsubscript{Ga} PL quenching is observed due to thermal emission of holes from the shallow state to the valence band.

To summarize the above results, in the case of bi-stable acceptors, i.e., the two states have nearly the same energies, such as for Mg\textsubscript{Ga} and Zn\textsubscript{Ga}, low-temperature PL intensity is determined by the nonradiative hole capture rate. This is in accord with the findings of Ref. \cite{Demchenko_Be_PRL}, where the PL intensity ratio of deep and shallow states was suggested to be determined by the respective hole capture coefficient ratio. If the two states have energies that are different by $\sim$10 meV, low temperature (i.e. at $T\sim$15 K) PL is also determined by the barrier between the two states. If this barrier is high for a thermal jump from the shallow to the deep state, then the shallow state will dominate the PL. 

\begin{figure}[b]
\vspace{0.1in}
\includegraphics[scale=0.32]{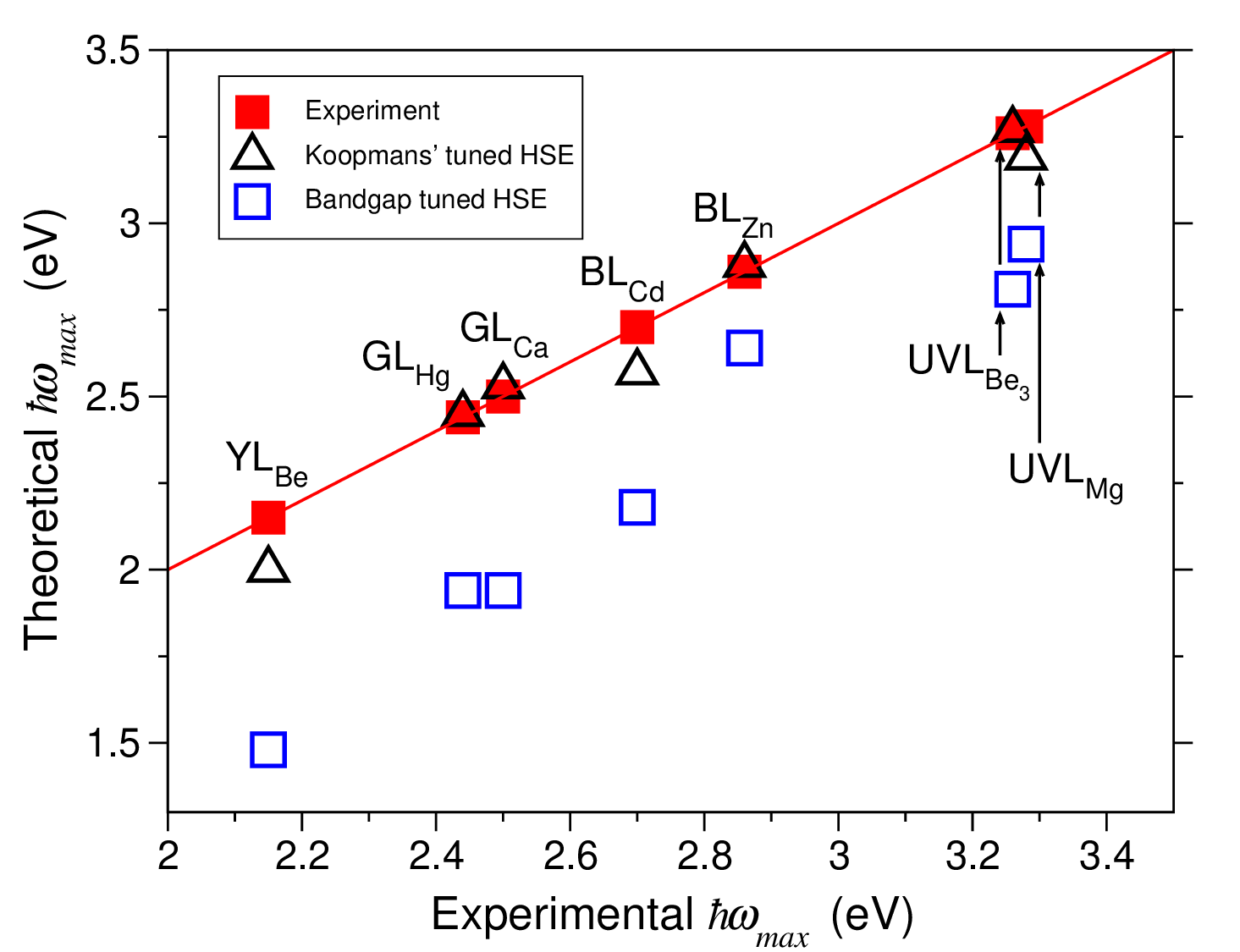}
\caption{\label{fig:theory-vs-experiment-Final} Comparison of theoretical results for PL maxima calculated using Koopmans tuned HSE and those obtained using the bandgap tuned HSE with the experiment.}
\end{figure}

\begin{table*}
\caption{\label{tab:Summary} Summary of HSE calculated properties of cation site acceptors in GaN compared to the experiment. Acceptor transition level 0/- (eV) above the VBM, the energy of the ZPL $E_0$ (eV), the energy of the PL maximum $\hbar \omega_{max}$ (eV), Franck-Condon shift $\Delta_{FC}$ (eV), the excited state vibrational level $\hbar \Omega_e$ (meV) and the Huang-Rhys factor $S_e$, and the ground state vibrational level $\hbar \Omega_g$ (meV) and the Huang-Rhys factor $S_g$.}
\begin{ruledtabular}
\begin{tabular} {c|cccccccc|cccccccc}
 &\multicolumn{8}{c}{Theory}&\multicolumn{8}{c}{Experiment}\\
 \hline
 PL & 0/- & $E_0$ & $\hbar \omega_{max}$ & $\Delta_{FC}$ & $\hbar \Omega_e$ & $S_e$ & $\hbar \Omega_g$ & $S_g$ & 0/- & $E_0$ & $\hbar \omega_{max}$  & $\Delta_{FC}$ & $\hbar \Omega_e$ & $S_e$ & $\hbar \Omega_g$ & $S_g$ \\
 \hline
 UVL\textsubscript{Be3} & 0.18 & 3.32 & 3.27 & 0.05 & 43 & 1 & 43 & 1 & 0.24 & 3.26 & 3.26 & 0 & - & $<1$ & 92 & - \\

YL\textsubscript{Be} & 0.34 & 3.16 & 2.0 & 1.17 & 17 & 17 & 34 & 34 & 0.38 & 3.12 & 2.15 & 1.0 & 38 & 22 & - & - \\

UVL\textsubscript{Mg} & 0.23 & 3.27 & 3.19 & 0.085 & 40 & 2 & 40 & 2 & 0.22 & 3.28 & 3.28 & 0 & - & 0.4 & 92 & - \\

BL\textsubscript{Zn} & 0.34 & 3.16 & 2.88 & 0.28 & 30 & 4.7 & 41.5 & 6.7 & 0.41 & 3.10 & 2.86 & 0.24 & 43 & 3.2 & 36, 92 & - \\

GL\textsubscript{Ca} & 0.42 & 3.08 & 2.53 & 0.55 & 25 & 11.6 & 34 & 16 & 0.5 & 3.0 & 2.5 & 0.5 & 41 & 8.5 & - & - \\

BL\textsubscript{Cd} & 0.5 & 3.0 & 2.57 & 0.43 & 31 & 7.5 & 42 & 10 & 0.55 & 2.95 & 2.7 & 0.26 & 53 & 4 & - & - \\

GL\textsubscript{Hg} & 0.67 & 2.83 & 2.45 & 0.38 & 25 & 10 & 31 & 12 & 0.77 & 2.73 & 2.44 & 0.29 & 53 & 5 & - & - \\

\end{tabular}
\end{ruledtabular}
\end{table*}

However, if the dual nature acceptor is not bi-stable, and one state is the ground state while another is metastable, the PL intensities at different temperatures depend on potential barriers between the states. The shallow state is still capturing most holes, but if the barrier from shallow to deep state is low for a thermal jump, the hole efficiently transfers to the ground state and the latter will dominate the PL. At the increased temperatures the hole is ejected back to the shallow state and both states produce PL. This is the situation observed for the Be\textsubscript{Ga} acceptor, where at low temperatures YL\textsubscript{Be} is observed, and at $T>140$ K, the UVL\textsubscript{Be3} is activated.

\section{HSE vs experiment}

It is illustrative to quantify whether the Koopmans tuning of the HSE hybrid functional is bringing theory closer to reproducing the experiment. 
Figure \ref{fig:theory-vs-experiment-Final} shows the comparison of the PL maxima obtained in this work using Koopmans tuned HSE with a commonly used bandgap-tuned HSE hybrid functional. The results are plotted against the experimental PL maxima. 
Note that Fig. \ref{fig:theory-vs-exp} shows a comparison between experiment and theory across different works, where different authors use different HSE parametrizations, supercell sizes, error correction schemes, \textbf{k}-point meshes, etc. 
Here, in order to compare the commonly used bandgap HSE tuning approach with the Koopmans tuning approach presented here, we recompute the results for the bandgap-tuned HSE using the same calculation setup (300 atom supercells, 500 eV energy cut-off, $\Gamma$-point only). 
The differences between the two theoretical approaches are the following. For Koopmans tuned HSE, the functional parameters were adjusted for each defect (see Table \ref{Table:HSE-tunings}). Errors due to periodic boundary conditions were corrected using the extrapolation scheme for both charged and neutral defects during the tuning procedure and for transition energy calculations. 
For the bandgap-tuned HSE, the same parameterization ($\alpha=0.31$, $\mu=0.2$ \AA$^{-1}$) was applied to all acceptors, and the artificial electrostatic interactions were corrected using the commonly used correction scheme \cite{FNV_1, FNV_2}, with no correction for defect state delocalization.

As shown in Fig. \ref{fig:theory-vs-experiment-Final} the bandgap-tuned HSE functional underestimates calculated PL maxima by 0.2-0.7 eV. This is the result of the overestimated thermodynamic transition levels and Franck-Condon shifts for all acceptors considered here. This, in turn, is a consequence of the bandgap-tuned HSE functional containing an excess of exact exchange energy in defect state orbitals leading to the deviation from the Koopmans' theorem and shifting acceptor transition levels deeper into the bandgap. In addition, this commonly used approach does not account for the wavefunction delocalization error, which leads to larger electrostatic corrections for negatively charged defects and no corrections for neutral defects. 
In contrast, an extrapolation-based correction scheme accounts for both electrostatics and delocalization, yielding HSE parametrizations that essentially produce self-interaction free defect orbitals. This significantly improves the transition energies of acceptors in GaN, albeit at the expense of an underestimated bandgap.

Table \ref{tab:Summary} shows the summary of theoretical and experimental optical and vibrational properties of cation-site acceptors in GaN. Overall, thermodynamic transition levels, PL maxima, and Frack-Condon shifts are in good agreement with the experiment. On the other hand, the defect PL band shapes, as computed from the HSE configuration coordinate diagrams along the direction between the neutral and negative charge states of acceptors, are less well reproduced. 
 At the same time, good fits can be obtained for all PL bands using the 1D configuration diagram model for an effective vibrational mode. This indicates that the 1D model is capable of describing the observed PL spectra. An alternative method is needed for finding the effective vibrational mode, which determines the PL band shape and energies of phonon replicas.

\section{Conclusions}

A systematic comparison between theory and experiment is presented for cation site acceptors in GaN. An extrapolation scheme is used to correct for errors arising from the use of the periodic boundary conditions. Two primary sources of error are the artificial electrostatic interaction between periodic images and defect state wavefunction delocalization in periodic supercells. These errors in total energy have opposite signs and partially cancel each other. 
Surprisingly, the two errors can be comparable in magnitude, exhibiting a delocalization error for some defects for which defect state charge density appears to be fully localized. This partial error cancellation leads to significantly lower than commonly used values of error correction for charged acceptors and nonzero corrections for some nominally neutral acceptors. 
Using this approach to error correction, the HSE hybrid functional was tuned to fulfill the generalized Koopmans' condition separately for all involved defect state orbitals. 
The Koopmans tuned HSE hybrid functional was used to calculate defect transition energies, PL band shapes, and acceptor vibrational parameters to compare them with low-temperature PL spectra from GaN samples doped with respective elements. 
The results show that this approach significantly improves the agreement between theory and experiment for all acceptors considered here. Calculated transition levels, PL maxima, and ZPL values are in very good agreement with the experiment. PL band shapes calculated using the 1-D configuration coordinate model and HSE transition energies exhibit more pronounced deviations from the experiment. In a 1-D model, the band shape is determined principally from the values of the ZPL and the excited state Huang-Rhys factor $S_e$. While ZPLs are well reproduced, the values of $S_e$ obtained from HSE show significant differences compared to the experiment. 
This is likely because a different effective vibrational mode determines the PL band shape rather than the commonly assumed mode along the direction connecting the two charge states of the defect.

\begin{acknowledgments}
The work at VCU and SUNY was supported by the National
Science Foundation under Grants No. DMR-1904861
and No. DMR-1905186, respectively. The calculations were performed at the VCU High Performance Research Computing (HPRC) core facility. 
\end{acknowledgments}


\providecommand{\noopsort}[1]{}\providecommand{\singleletter}[1]{#1}%

\end{document}